\documentclass[a4paper]{ar-1col}
\usepackage{rotating}
\usepackage{graphicx}
\usepackage{natbib}
\usepackage{amsmath}
\usepackage{upgreek}
\usepackage{textcomp}
\usepackage{wasysym}
\usepackage[version=3]{mhchem} % Package for chemical equation typesetting
\usepackage{totcount}
\newtotcounter{citnum} %From the package documentation
\def\oldbibitem{} \let\oldbibitem=\bibitem
\def\bibitem{\stepcounter{citnum}\oldbibitem}

\jname{Annual Review of Earth \& Planetary Sciences}
\jvol{44}
\jyear{2016}
\doi{10.1146/xxxxxxx}

\begin{document}
% Page heads
\markboth{Robin Wordsworth}{The Climate of Early Mars}

% Title
\title{The Climate of Early Mars}

% Author/affiliation
\author{Robin D. Wordsworth$^{1,2}$
\affil{$^1$Harvard Paulson School of Engineering and Applied Sciences, Harvard University, Cambridge, MA 02140, USA}
\affil{$^2$Department of Earth and Planetary Sciences, Harvard University, Cambridge, MA 02140, USA}}

\begin{abstract}
The nature of the early Martian climate is one of the major unanswered questions of planetary science. Key challenges remain, but a new wave of orbital and in situ observations and improvements in climate modeling have led to significant advances over the last decade. Multiple lines of geologic evidence now point to an episodically warm surface during the late Noachian and early Hesperian periods 3-4~Ga. The low solar flux received by Mars in its first billion years and inefficiency of plausible greenhouse gases such as \ce{CO2} means that the steady-state early Martian climate was likely cold. A denser \ce{CO2} atmosphere would have caused adiabatic cooling of the surface and hence migration of water ice to the higher altitude equatorial and southern regions of the planet. Transient warming caused melting of snow and ice deposits and a temporarily active hydrological cycle, leading to erosion of the valley networks and other fluvial features.  Precise details of the warming mechanisms remain unclear, but impacts, volcanism and orbital forcing all likely played an important role. The lack of evidence for glaciation across much of Mars' ancient terrain suggests the late Noachian surface water inventory was not sufficient to sustain a northern ocean. While mainly inhospitable on the surface, early Mars may nonetheless have presented significant opportunities for the development of microbial life. 
\end{abstract}

\begin{keywords}
mars, paleoclimate, atmospheric evolution, faint young Sun, astrobiology
\end{keywords}

\maketitle

\section{INTRODUCTION}

With the exception of Earth, Mars is the Solar System's best-studied planet. Since the first \emph{Mariner} flybys in the 1960s, Mars has been successfully observed by a total of 11 orbiters and 7 landers, four of them rovers. Combined with ongoing observations from Earth, this has allowed a uniquely comprehensive description of the Martian atmosphere and surface. However, despite the wealth of data obtained, fundamental mysteries about Mars' evolution remain. The biggest mystery of all is the nature of the early climate: 3-4~Ga Mars should have been freezing cold, but there is nonetheless abundant evidence that liquid water flowed across its surface.

Unlike Earth, Mars lacks plate tectonics, global oceans and a biosphere\footnote{While the possibility of life on Mars today still cannot be entirely ruled out, the absence of a biosphere sufficient to modify the surface substantially is clear.}. One of the happy outcomes of this is that its ancient crust is incredibly well-preserved, allowing a window to epochs as early as 3-4~Ga across large regions of the surface \citep{Nimmo2005}. This antiquity is hard to imagine from a terrestrial perspective. By way of comparison, we can imagine the advantages to Precambrian geology if most of Asia consisted of lightly altered terrain from the early Archean, when life was first emerging on Earth. Mars provides us a glimpse of conditions during the earliest stages of the Solar System on a body that had an atmosphere, at least episodic surface liquid water, and in some locales, surface chemistry conducive to the survival of microbial life \citep[e.g., ][]{Grotzinger2014}. 

There are many motivations for studying the early climate of Mars. The first is simply that it is a fundamentally interesting unsolved problem in planetary science. Another major motivation is  astrobiological --- if we can understand how the Martian climate evolved, we will have a better understanding of whether life could have ever flourished, and where to look for it if it did. 
Studying Mars also has the potential to inform us about the evolution of our own planet, because many of the processes thought to be significant to climate on early Mars (e.g. volcanism, impacts) have also been of major importance on Earth. Finally, in this era of exoplanet science, Mars also represents a test case that can inform us about the climates of small rocky planets in general.

The aim of this review is to provide a general introduction to the latest research on the early Martian climate. We begin by discussing highlights of the geologic evidence for an altered climate on early Mars, focusing on the extent to which the observations are consistent with episodic warming vs. a steady-state warm and wet climate. Next, we discuss the external boundary conditions on the early climate (namely the solar flux and Martian orbital parameters) and the constraints on the early atmospheric pressure. We also review previous one-dimensional radiative-convective modeling of the effects of key processes (atmospheric composition, meteorite impacts and volcanism) on surface temperature. Finally, we discuss recent three-dimensional climate modeling of early Mars by a number of groups that has increased our understanding of cloud and aerosol processes and the nature of the early water cycle. It is argued that future progress will require an integrated approach, where three-dimensional climate models are compared with the geologic evidence on both global and regional scales.

\section{GEOLOGIC EVIDENCE FOR LIQUID WATER ON EARLY MARS}\label{sec:geo_evid}

With a few important exceptions, all our current information on the early martian climate comes from surface geology. Martian geologic data is derived from a combination of passive and active orbital remote sensing and \emph{in situ} analysis. The oldest and best studied aspect of Martian geology is the surface geomorphology. In recent years, the geomorphic data have been supplemented by global maps of surface mineralogy derived from orbiters and detailed \emph{in situ} studies of several specific regions by the NASA rover missions. 

\begin{sidewaysfigure}
\includegraphics[height=3.5in]{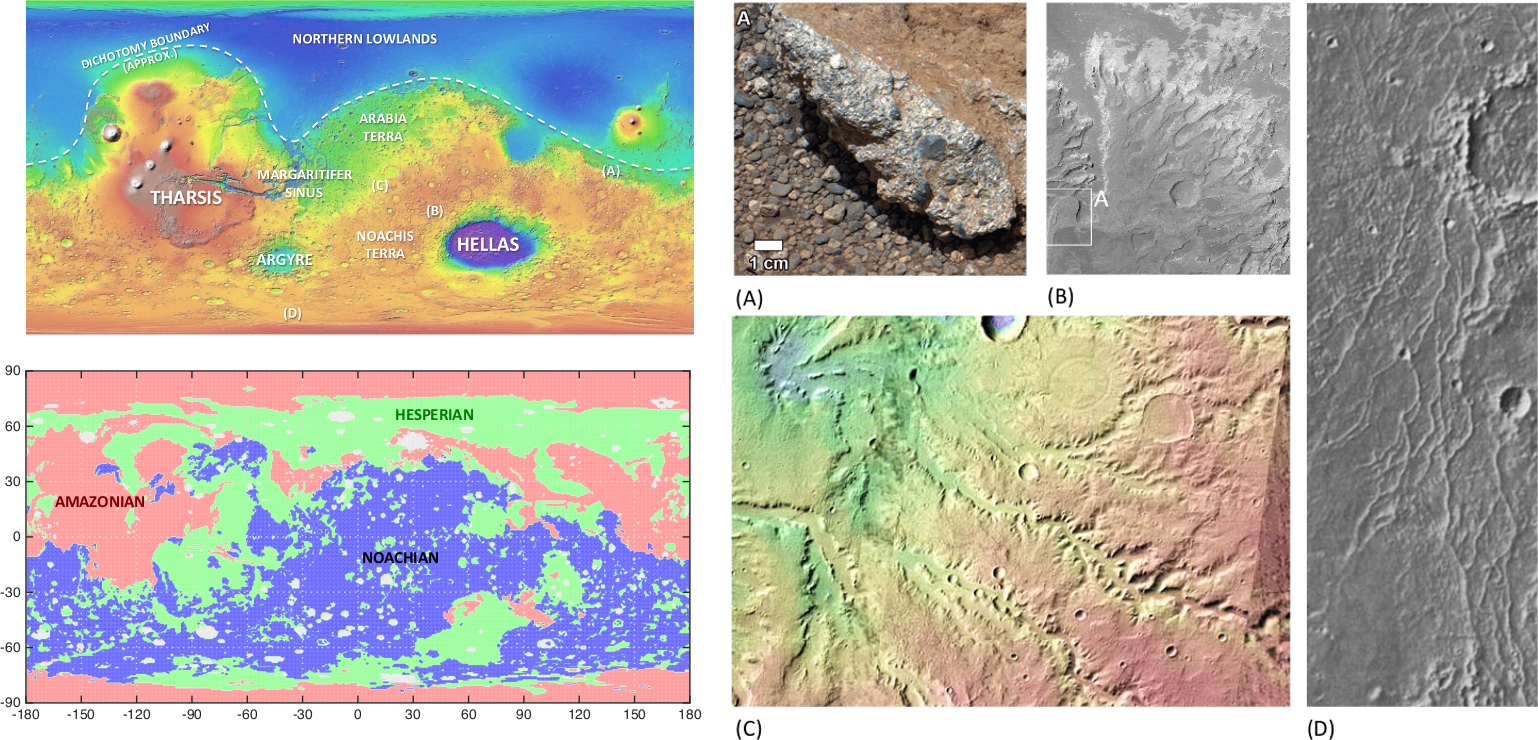}
\caption{(left) Contour plot of the present-day Martian topography from MOLA data. Major features of the terrain are indicated. (bottom left) Spatial distribution of the three main terrain types on the Martian surface [data from \cite{Tanaka1986,Scott1986,Tanaka1987}]. (right) Highlights of the geomorphological evidence for an altered climate in the Noachian and Hesperian. (A) Fluvial conglomerates observed \emph{in situ} by the Curiosity rover at Gale Crater [from \cite{Williams2013}; reprinted with permission from AAAS]. (B) Deltaic lake deposits in Eberswalde crater [from \cite{Malin2003}; reprinted with permission from AAAS]. (C) Valley networks in Paran\'a Vallis [from \cite{Howard2005}]. (D) Sinuous ridges in the Dorsa Argentea Formation interpreted as glacial eskers [from \cite{HeadPratt2001}].}\label{fig:mars_overview}
\end{sidewaysfigure}

Figure~\ref{fig:mars_overview} (top left) summarizes the basic features of the Martian surface. The four most important large-scale features are the north-south dichotomy, the Tharsis bulge, and the Hellas and Argyre impact craters.  Because Martian topography plays a major role in the planet's climate and hydrological cycle, understanding when these features formed is vital. The relative ages of surface features and regions on Mars can be assessed via analysis of local crater size-frequency distributions (crater statistics) \citep{Tanaka1986,Tanaka2014}. Absent geochronology data, \emph{absolute} dating of Martian surface units relies on impactor flux models and hence is subject to considerable uncertainty.

\begin{marginnote}
\entry{Tharsis bulge}{A large region of elevated terrain that dominates the equatorial topography of Mars.}
\end{marginnote}

\begin{figure}
\includegraphics[width=4.5in]{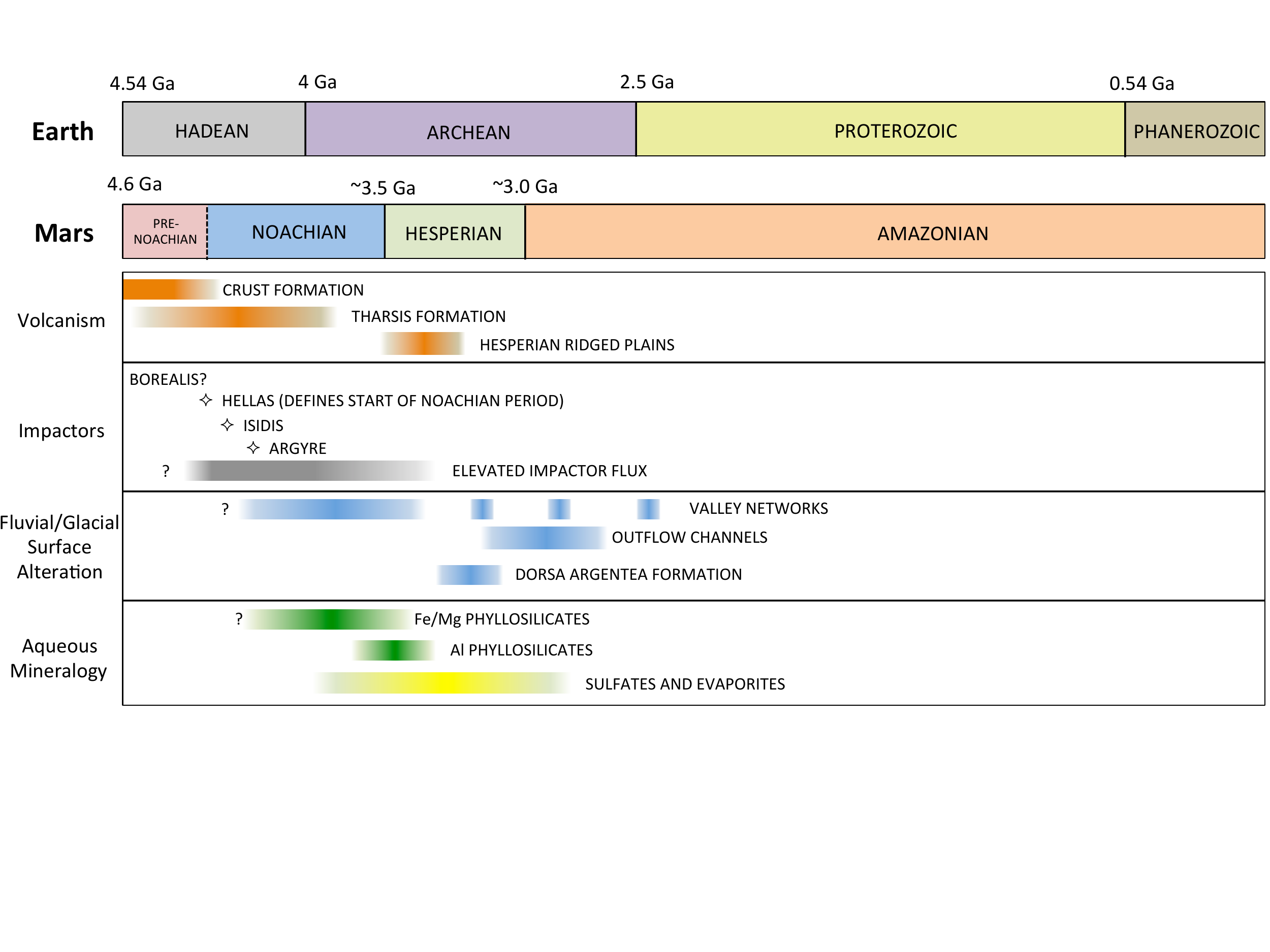}
\caption{Timeline of major events in Mars history, with the geologic eons of Earth displayed above. In general, the absolute timing of events on Mars is subject to considerable uncertainty, but the sequencing is much more robust. Question marks indicate cases where processes could also have occurred earlier but the geologic record is obscured by subsequent events. Based on data from \cite{Werner2011,Fassett2011,Ehlmann2011} and \cite{HeadPratt2001}.}\label{fig:timeline}
\end{figure}

It is standard to categorize Martian terrain into three time periods (Fig.~\ref{fig:timeline}): the most modern Amazonian ($\sim 0-3.0$~Ga), which is associated with hyperarid, oxidising surface conditions and minimal weathering; the Hesperian ($\sim 3.0-3.5$~Ga), which contains evidence of extensive volcanism and catastrophic flooding, and the ancient Noachian ($\sim 3.5-4.1$~Ga), when alteration of the Martian surface by water was greatest [\cite{Werner2011}; also see Fig.~\ref{fig:mars_overview}]. The vast majority of Noachian units are found in the heavily cratered south. The northern hemisphere is dominated by smooth plains that probably result from lava outflow and are dated to the Hesperian and Amazonian \citep{Tanaka1986}. The Noachian period is defined by the Noachis Terra region, which translates evocatively as ``Land of Noah". It contains the clearest evidence for an altered early climate and is the primary focus of this article.

All of the largest scale topographic features on Mars formed during or before the Noachian. The north-south dichotomy is the most ancient surface feature, followed by Hellas and Argyre. The Hellas impact is commonly taken to represent the pre-Noachian/Noachian boundary \citep{Nimmo2005,Fassett2011}. Although substantial resurfacing and formation of the Tharsis Montes shield volcanoes was ongoing during the Hesperian and Amazonian, the formation of Tharsis likely began early and was mainly complete by the late Noachian \citep{Phillips2001,Carr2010,Fassett2011}. To a first approximation, the large-scale topography of Mars in the late Noachian was therefore probably similar to that today.

\begin{marginnote}
\entry{Hellas}{Largest confirmed impact basin on Mars; defines the start of the Noachian period}
\end{marginnote}

\subsection{Geomorphology}
\subsubsection{Valley Networks}

The valley networks are the single most important piece of evidence in favor of a radically different climate on early Mars. Like many drainage basins on Earth and in contrast with the later Hesperian-period outflow channels, Martian valley networks are dendritic (branching) with tributaries that begin near the peaks of topographic divides. This geomorphology strongly suggests an origin due to a hydrological cycle driven by precipitation (as rain or snow) \citep{Craddock2002,Mangold2004,Stepinski2005,Barnhart2009,Hynek2010,Matsubara2013} rather than e.g., groundwater sapping \citep{Squyres1994} or basal melting of thick icesheets \citep{Carr2003}. 

Valley networks are rare on Hesperian and Amazonian terrain but common on Noachian terrain, where they are predominantly seen at equatorial latitudes between $60^\circ$~S and $10^\circ$~N 
\citep{Milton1973,Carr1996,Hynek2010}. The largest networks are huge, extending thousands of kilometers over the surface in some cases \citep{Howard2005,Hoke2011}. Landform evolution models suggest minimum formation timescales for the valley networks of $10^5$ to $10^7$ years under climate conditions appropriate to arid regions on Earth \citep{Barnhart2009,Hoke2011}.

\begin{marginnote}
\entry{dendritic valley networks}{branching networks of channels carved into the ancient Martian crust that share many similarities with drainage basins on Earth}
\end{marginnote}

\subsubsection{Crater Lakes}

When liquid water carves valley networks on a heavily cratered terrain, ponding and lake formation inside craters is a natural outcome. For sufficiently high flow rates, crater lakes will breach their rims, forming open lakes that are integrated in a larger hydrological network. Both the ratio of watershed area to lake area (drainage ratio) for each lake and the ratio of open to closed basin crater lakes in total give important clues as to the nature of the Noachian water cycle. In general, low drainage ratios and a large number of open crater lakes indicate high precipitation rates and a wet climate \citep{Fassett2008,Barnhart2009}.

As might be expected, analysis of the Noachian southern highlands has revealed abundant evidence for crater lakes interlinked with the valley networks \citep{Cabrol1999,Fassett2008}. However, closed-basin lakes greatly outnumber open-basin lakes. This suggests that a very wet climate or periodic catastrophic deluges due to e.g. impact-driven steam greenhouses (see Section~\ref{subsec:episodic}) were \emph{not} responsible for their formation \citep{Irwin2005,Barnhart2009}. Open-basin lake drainage ratios strongly vary with location, with wetter formation conditions indicated in Arabia Terra and north of Hellas (Terra Sabaea) \citep{Fassett2008}. 

Striking evidence of \emph{in situ} fluvial erosion was found by NASA's Curiosity rover in the form of conglomerate outcrops at Gale Crater \citep{Williams2013}. Morphologically, the conglomerates are remarkably similar to sediment deposits found on Earth (see Fig.~\ref{fig:mars_overview}). However, chemical analysis of the outcrops suggested low chemical alteration of the material by water \citep{Williams2013}. Indeed, global analysis of aqueous alteration products on the Martian surface suggests a predominance of juvenile or weakly modified minerals \citep{Tosca2009}. In addition, most of the minerals in open-basin crater lakes observed from orbit lack evidence of strong \emph{in situ} chemical alteration \citep{Goudge2012}. This suggests that the flows responsible for eroding the late Noachian and Hesperian surface were relatively short-lived.

\subsubsection{Northern Ocean} The most evocative (and controversial) claim to have come out of geomorphic studies of the ancient surface is that Mars once possessed a northern ocean of liquid water. The original argument proposed to support this is that various geologic contacts in the northern plains resemble ancient shorelines \citep{Parker1993,Head1999}. Several of the putative shorelines show vertical variations of several kilometers, which is inconsistent with a fluid in hydrostatic equilibrium, although it has been argued that true polar wander could have caused surface deformation sufficient to explain this \citep{Perron2007}. More critically, much of the shoreline evidence was found to be ambiguous in subsequent high-resolution imaging studies \citep{Malin1999}. 

More recently, it has been argued that many delta-like deposits, which are assumed to be of fluvial origin, follow an isostatic line at -2.54~km from the datum across the surface \citep{diAchille2010}. If this line does represent a Noachian ocean shoreline, the implications are a) that the earlier proposed shorelines are incorrect and Martian topography was not modified by a true polar wander event and b) Mars once had a global equivalent layer (GEL) of around 550~m of surface water. The extent to which a warm and wet scenario for early Mars with a northern ocean fits the climate constraints and the other geologic evidence is a major focus of the rest of this article.

\begin{marginnote}
\entry{GEL}{global equivalent layer (of water) averaged across the planet's surface}
\end{marginnote}

\subsubsection{Glaciation} The evidence for an at least episodically warmer early Martian climate is not limited to fluvial landforms. At the south pole, the Dorsa Argentea Formation (DAF), a geologic unit dated to the mid-Hesperian, contains a range of features suggestive of glaciation, including sinuous ridges interpreted as eskers (Fig.~\ref{fig:mars_overview}) and pitted regions that may have been caused by basal melting of a thick ice sheet \citep{Howard1981,HeadPratt2001}. Around the Argyre and Hellas basins, further glacial landforms such as eskers, lobate debris aprons and possible moraines and cirques are observed \citep{Kargel1992}. Dynamic icesheet modeling \citep{Fastook2012} suggests that polar surface temperature increases of 25-50~K from Amazonian (modern) values are required before wet-based glaciation of the DAF could occur, again suggesting episodically warmer climate conditions on early Mars. Interestingly, however there is comparatively little evidence for glacial alteration of the surface on Noachian or Hesperian terrain at more equatorial latitudes. This is an important issue that we return to in Section~\ref{sec:watercycle}.

\begin{marginnote}
\entry{DAF}{Dorsa Argentea Formation: mid-Hesperian geologic unit at Mars' south pole interpreted as the remains of a large water ice cap.}
\end{marginnote}

\begin{marginnote}
\entry{lobate debris apron}{Distinctive volatile-rich, convex Martian landform analogous to a terrestrial rock glacier.}
\end{marginnote}

\subsection{Geochemistry}

The morphological evidence for liquid water on early Mars, which has been observed in increasing detail from the 1960s onwards, has been complemented over the last 10-15 years by a new array of geochemical observations from orbit and rover missions. Iron and magnesium rich phyllosilicates (clays) are found extensively over Noachian terrain \citep{Poulet2005,Bibring2006,Mustard2008,Murchie2009,Carter2010}. To form, these minerals require the presence of liquid water and near-neutral-pH conditions. Other aqueous minerals such as sulphates, chlorides and silicas are found in more localized regions of the Noachian and Hesperian crust \citep{Gendrin2005,Osterloo2010,Carter2013,Ehlmann2014}. 

If the phyllosilicates mainly formed on the surface, this would represent evidence in favor of a warm and wet early Martian climate \citep{Poulet2005,Bibring2006}. Recently, however, it has been argued that in most cases their mineralogy may best represent subsurface formation in geothermally heated, water-poor systems \citep{Ehlmann2011}. Many of the sedimentary phyllosilicates observed on the Martian surface today may not have formed \emph{in situ}, but were instead transported to their current locations by later erosion of the crust. 

At several sites, Al-rich clays such as kaolinite are observed alongside or overlying Fe/Mg clays \citep{Poulet2005,Wray2008,Ehlmann2009,Carter2015}. On Earth, Al-rich clays overlying Fe/Mg clay in a stratigraphic section is a common feature of wet environments, because iron and magnesium cations from the original minerals are preferentially leached (flushed) downwards from the topmost layer by water from rain or snowmelt. This is one possible explanation for the presence of Al-rich clays on Mars \citep{Carter2015}. Another interpretation is more acidic and oxidizing local alteration conditions in a mainly cold climate, as suggested by the presence of the sulfate mineral jarosite adjacent to Al clays in several regions \citep{Ehlmann2015_LPSC}.

Sulfate deposits on Mars are particularly interesting because they require a source (most likely volcanic) of sulfur and in some cases indicate acidic and/or saline formation conditions. Sulfates appear primarily, but not exclusively, on Hesperian terrain and may be associated with the volcanic activity that formed the basaltic ridged plains \citep{Head2002,Bibring2006}. The link between the sulfates, volcanism and possible changes in the early climate due to sulfur dioxide (\ce{SO2}) and hydrogen sulfide (\ce{H2S}) is discussed in Section~\ref{subsec:sulf}.

\begin{marginnote}
\entry{CRISM}{Compact Reconnaissance Imaging Spectrometer for Mars: a visible-infrared spectrometer onboard the Mars Reconnaissance Orbiter}
\end{marginnote}

One mineral, carbonate, is conspicuous by its \emph{low} abundance on the Martian surface \citep{Niles2013,Ehlmann2014}. This is important, because surface carbonate formation should be very efficient on a warm and wet planet with a basaltic crust and \ce{CO2}-rich atmosphere \citep{Pollack1987}. Carbonate formation could have been suppressed by acidic surface conditions caused by dissolution of \ce{SO2} in water \citep{Bullock2007,Halevy2007}. However, globally acidic warm and wet conditions are difficult to justify in the presence of a strongly mafic basaltic regolith, which should effectively buffer pH, just like the basaltic seafloor on Earth \citep{Niles2013}. Hence the absence of surface carbonates is a strong indication that early Mars was either only episodically warm, or very dry. Interestingly, carbonates \emph{have} been discovered in outcrops in the  Nili Fossae region by CRISM \citep{Ehlmann2008} and in Gusev crater by the Spirit rover \citep{Morris2010}. They are also seen in some regions where deep crustal material has been excavated by impacts \citep{Michalski2010}. This indicates that regardless of the early surface conditions, the deep crust may still have been a major sink for atmospheric \ce{CO2} over time (Section~\ref{subsec:howthick}).

\section{FAINT YOUNG SUN, COLD YOUNG PLANET?}

The geologic record is unanimous: liquid water substantially modified Mars' surface during the late Noachian. The surface processes that could have created this water, however, are far from obvious. 
Two basic facts conspire to make warming early Mars a fiendish challenge: the Martian orbit and the faintness of the young Sun. 

With a semi-major axis of 1.524~AU, Mars receives around 43\% of the solar energy that Earth does. The rapid dissipation of the nebula during terrestrial planet formation and lack of major configurational changes in the Solar System since the late heavy bombardment means that Mars' orbital semi-major axis cannot have changed significantly since the late Noachian. Mars' orbital eccentricity and obliquity evolve chaotically on long timescales, however, and have probably varied over ranges of 0 -- 0.125 and 10$^\circ$ -- 60$^\circ$, respectively \citep{Laskar2004}.  Although this has little effect on the net annual solar flux, it still has important implications for the early climate, because the time-varying insolation pattern is a key determinant of peak summertime temperatures and hence the long-term transport and melting of water ice. 

The early Sun was less luminous than today because hydrogen burning increases the mean molar mass of the core, causing it to contract and heat up. The rate of fusion is strongly dependent on temperature, so this in turn increases a main sequence star's luminosity over time. This fundamental outcome of stellar physics is supported by detailed solar models and observations of many nearby stars. As a result, the Sun's luminosity 3.8~Ga was approximately 75\% of its present-day value \citep{Gough1981}. One possible way to avoid this outcome is if the Sun shed large amounts of its mass early on [over 2\% in the first 2~Gyr; \cite{Minton2007}]. While possible, this is unlikely, because such high mass loss is not observed in any nearby G or K class stars \citep{Minton2007}. 
The idea that our Sun must be a unique and unusual star solely because Mars once had surface liquid water has not gained widespread acceptance.

If we accept the standard orbital and solar boundary conditions, the early Mars climate problem is now simply understood. Let us assume that in the late Noachian, Mars' received solar flux was $0.75\times 1366/1.524^2 = 441.1$~W~m$^{-2}$. If the planetary albedo were zero (i.e. every solar photon intersected by Mars was absorbed), the equilibrium temperature would then be {$T_e = [441.1 /4\sigma ]^{1/4}=210$~K} (here $\sigma$ is the Stefan-Boltzmann constant). This implies a \emph{minimum} greenhouse effect of around 65~K (around double that of present-day Earth or 9 times that of present-day Mars) to achieve even marginally warm and wet surface conditions. For more realistic planetary albedo estimates, the greenhouse effect required is tens of degrees greater still. 

\subsection{A Denser Early Atmosphere}\label{subsec:howthick}

One seemingly obvious way to invoke a more potent greenhouse effect on early Mars is via a denser atmosphere. But how thick could the early atmosphere have been? Estimating the total atmospheric pressure in the late Noachian requires consideration of the major sources (volcanic outgassing and impact delivery) and sinks (escape to space and incorporation of \ce{CO2} into the crust).

Carbon dioxide is generally assumed to have been the dominant constituent of the Martian atmosphere in the late Noachian, as it is today. The outgassing of \ce{CO2} into the Martian atmosphere with time is a function of the rate of volcanic activity and the chemical composition of the mantle \citep{Grott2011}. The rate of volcanism through the pre-Noachian and Noachian is not strongly constrained, but the majority of volatile outgassing in Mars' history almost certainly occurred in these periods \citep{Grott2011}.

Regarding the chemical composition, \ce{CO2} outgassing is strongly dependent in particular on the mantle oxygen fugacity ($f_\ce{O2}$). Analysis of  Martian meteorites suggests Mars has a more reducing mantle than Earth, with an $f_\ce{O2}$ value between the iron-w\"ustite and quartz-fayalite-magnetite buffers \citep{Wadhwa2001}. Given this, \cite{HirschmannWithers2008} estimate that between 70~mbar and 13~bar of \ce{CO2} could have been outgassed during the initial formation of the Martian crust, with the lower estimate for the most reducing conditions. During later events (such as the formation of the Tharsis bulge), they estimate that 40~mbar to 1.4~bars could have been outgassed.

\begin{marginnote}
\entry{XUV}{extreme ultraviolet}
\end{marginnote}

The escape rate of \ce{CO2} to space until the Hesperian is highly uncertain. Escape rates were highest just after Mars' formation, and the isotopic fractionation of noble gases in the atmosphere indicates the majority of the primordial atmosphere was lost very early \citep{Jakosky1997}. It has also been argued that all of the initially outgassed \ce{CO2} would have been rapidly lost to space by XUV-driven escape before the late Noachian \citep{Tian2010}. However, effective loss requires total dissociation of \ce{CO2} into its constituent atoms. The chemistry of this process has not been extensively studied and may be somewhat model-dependent \citep{Lammer2013}. Meteorite impacts during accretion remove \ce{CO2} but also deliver it, with the balance dependent on the model used. In contrast, escape processes occurring from the Hesperian onwards appear unambiguously ineffective: ion escape, plasma instability, sputtering and non-thermal processes combined could not have removed more than a few 100~mbar at most since 4~Ga \citep{Chassefiere2004,Lammer2013}.  Although further insights into these processes will be supplied by NASA's ongoing MAVEN mission, it currently appears that a dense late Noachian atmosphere  can only have been removed subsequently by surface processes.

The most efficient potential sink for atmospheric \ce{CO2} at the surface is carbonate formation.  As we have discussed, surface carbonates are rare on Mars. However, the discovery of carbonates in exhumed deep crust suggests that the possibility of a large subsurface reservoir cannot be discounted \citep{Michalski2010}. Recently, it has been argued based on orbital observations that this reservoir is unlikely to allow more than a 500~mbar \ce{CO2} atmosphere during the late Noachian \citep{Edwards2015}.  However, sequestration by hydrothermal circulation of \ce{CO2} in deep basaltic crust is a very poorly understood process even on Earth \citep[e.g.,][]{Brady1997}, so caution is still required when extrapolating the known carbonate reservoirs.  It will be hard to constrain the late Noachian carbon budget definitively until we can send missions to Mars (robotic or human) that are capable of drilling deep into the subsurface. 

Finally, one independent constraint on atmospheric pressure comes from the size distribution of craters on ancient terrain. In a thick atmosphere smaller impactors burn up before they reach the surface, so observation of the smallest craters leads to an upper limit on atmospheric pressure. Recent analysis of the Dorsa Aeolis region near Gale crater using this technique has led to an approximate upper limit of 0.9-3.8~bars on atmospheric pressure 3.6~Ga \citep{Kite2014}. 

To summarize, many aspects of Mars' atmospheric evolution are highly uncertain. It is likely that Mars had a thicker \ce{CO2} atmosphere in the late Noachian. This atmosphere could have been as dense as 1-2~bar, but likely no more than this. If the early atmosphere was denser than about 0.5~bar, it cannot have all escaped to space and the difference will now be buried in the deep crust as carbonate. Several recent studies have suggested that this reservoir may be small, but the observational search for carbonate deposits on Mars should continue, along with theoretical study of the interaction between atmospheric \ce{CO2} and pore water in deep Martian hydrothermal systems.

\subsection{The Failure of the \ce{CO2} Greenhouse}

Constraining the early atmospheric \ce{CO2} content is necessary to build a complete picture of the Noachian climate, but it is not sufficient. In a seminal paper, \cite{Kasting1991} demonstrated that regardless of the atmospheric pressure, a clear-sky \ce{CO2}-\ce{H2O} atmosphere alone could not have warmed early Mars. There are two reasons for this. First, \ce{CO2} is an efficient Rayleigh scatterer, so in large quantities it significantly raises the planetary albedo. In addition, \ce{CO2} condenses into clouds of dry ice at low temperatures. As surface pressure increases this leads to a shallower atmospheric lapse rate, reducing the greenhouse effect (Fig.~\ref{fig:TandP}). At high enough \ce{CO2} pressures, the atmosphere collapses on the surface completely. This conclusion, which was reached by Kasting using a one-dimensional clear-sky radiative-convective climate model, has recently been confirmed by three-dimensional climate models that include cloud effects \citep{Forget2013,Wordsworth2013a}. 

\begin{marginnote}
\entry{CIA}{collision-induced absorption}
\end{marginnote}
\begin{marginnote}
\entry{radiative-convective model}{Climate model that resolves atmospheric vertical structure and calculates radiative transfer accurately but lacks horizontal resolution.}
\end{marginnote}

\begin{figure}
\includegraphics[width=4.5in]{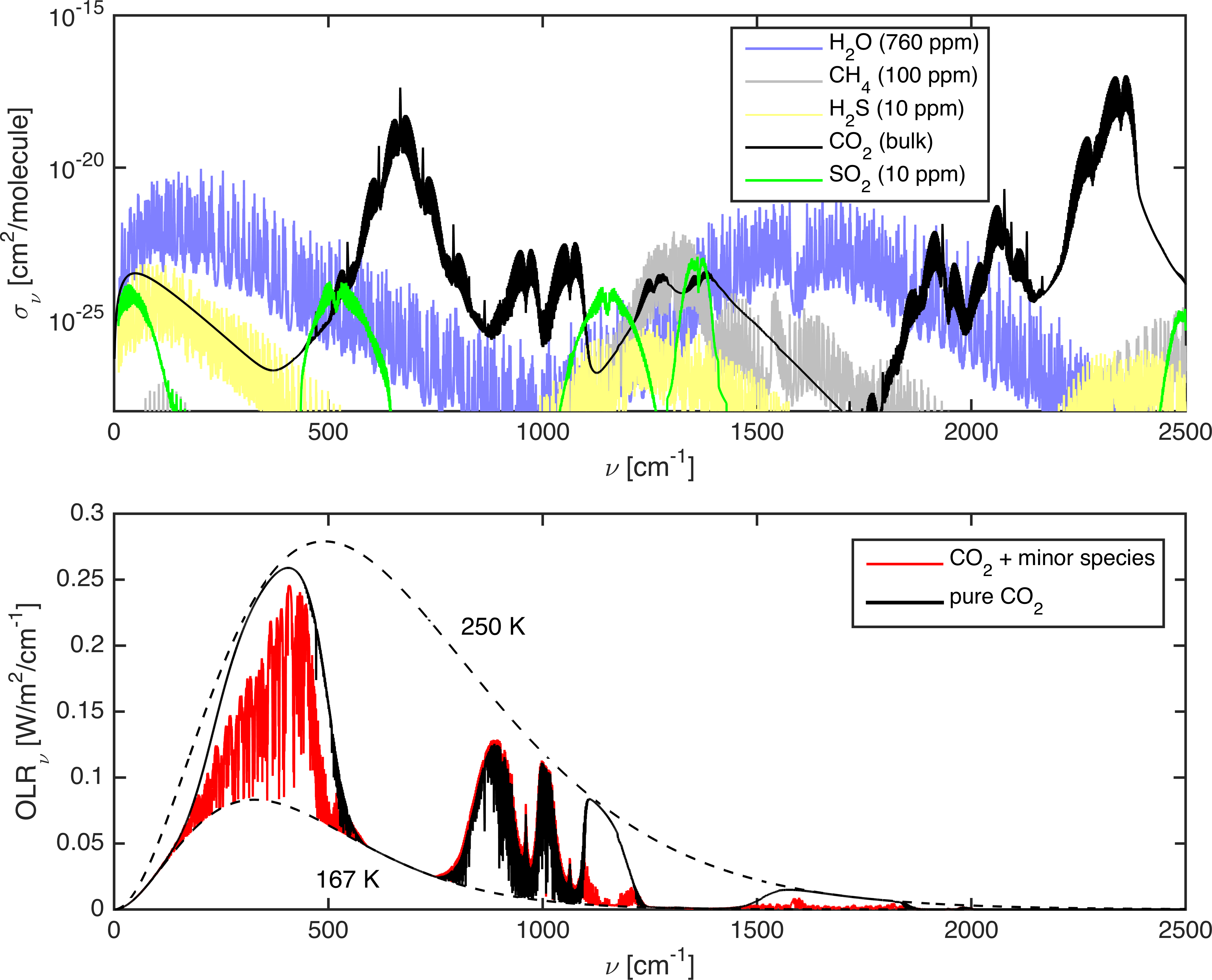}
\caption{The greenhouse effect on early Mars. (top) Absorption cross-sections per molecule of background gas vs. wavenumber at 1~bar and 250~K, for various greenhouse gases in the early Martian atmosphere, with the gas abundances given in the legend. Results were produced using the open-source software \emph{kspectrum}. (bottom) OLR vs. wavenumber from early Mars calculated using a line-by-line calculation assuming surface pressure of 1~bar, surface temperature of 250~K and a 167~K isothermal stratosphere. Blackbody emission at 250~K and 167~K is indicated by the dashed lines. The black line shows OLR for a pure \ce{CO2} atmosphere, while the red line shows OLR with all the additional greenhouse gases in the top plot included. Results were produced using the HITRAN 2012 database, the \cite{Clough1992} approach to solving the infrared radiative transfer equation and the GBB \ce{CO2} CIA parametrization from \cite{Wordsworth2010}. 
}\label{fig:spectra}
\end{figure}

\begin{figure}
\includegraphics[width=4.5in]{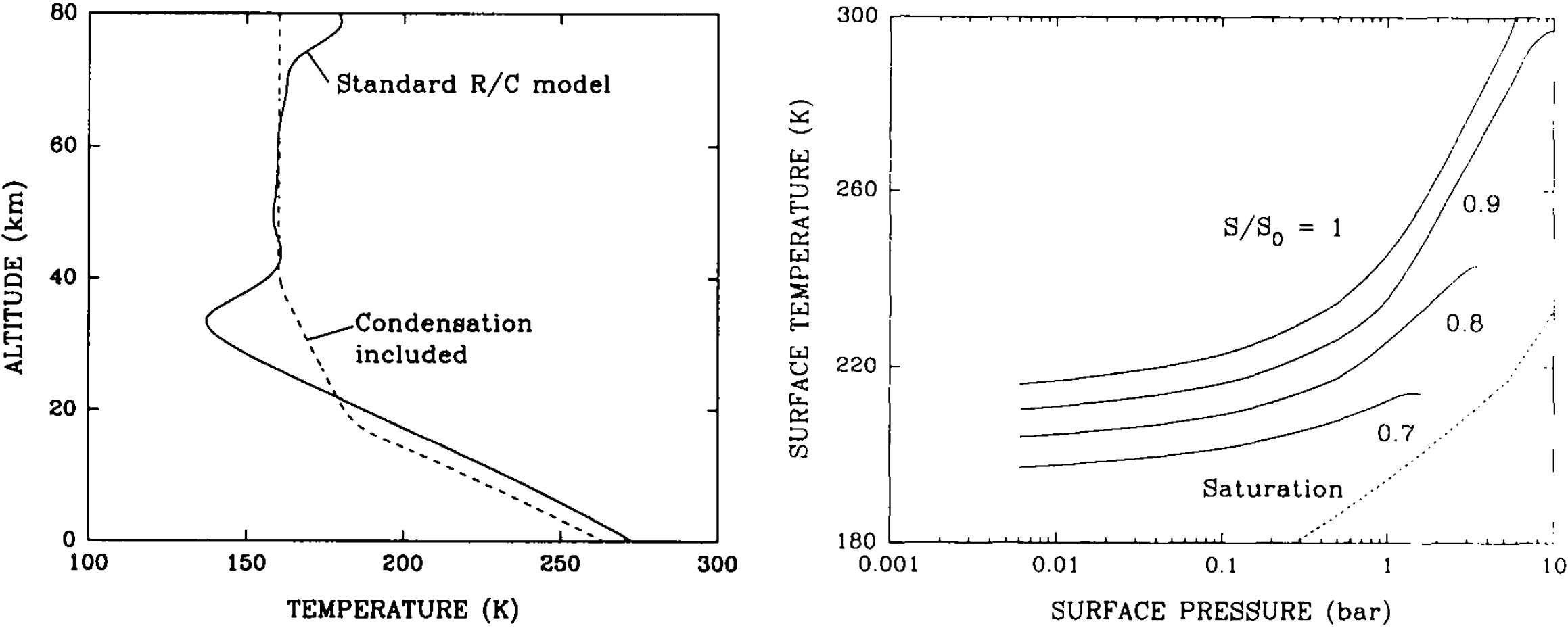}
\caption{The two figures that began the modern era of research on the early Martian climate [from \cite{Kasting1991}]. (left) Temperature-pressure profile for the early Martian atmosphere assuming a surface pressure of 2~bar. The dashed line shows the case where \ce{CO2} condensation is (correctly) included, leading to a weaker greenhouse effect. (right) Surface temperature vs. pressure produced from a clear-sky one-dimensional radiative-convective climate model for several values of solar luminosity relative to present day. The dashed line shows the saturation vapor pressure of \ce{CO2}. Note that these surface temperatures are now regarded as \emph{overestimates}, due to the problems in representation of the \ce{CO2} CIA described in \cite{Halevy2009} and \cite{Wordsworth2010}.}\label{fig:TandP}
\end{figure}

Although uncertainty remains, the infrared radiative effects of dense \ce{CO2}-dominated atmospheres are now fairly well understood. \ce{CO2} is opaque across important regions of the infrared because of direct vibrational-rotational absorption, particularly due to the $\nu_2$ 667~cm$^{-1}$ (15~$\upmu$m) bending mode and associated overtone bands. In dense atmospheres, \ce{CO2}, like most gases, also absorbs effectively through collision-induced absorption (CIA). CIA is a collective effect that involves the interaction of electromagnetic radiation with pairs (or larger numbers) of molecules. For \ce{CO2}, it occurs due to both induced dipole effects in the 0-250~cm$^{-1}$ region \citep{Gruszka1997}, and dimer effects between 1200 and 1500~cm$^{-1}$ \citep{Baranov2004} (see Fig.~\ref{fig:spectra}). Further complications arise from the fact that the sub-Lorentzian nature of absorption lines far from their centers must also be taken into account for accurate computation of \ce{CO2} absorption in climate models \citep{Perrin1989}. 

Figure~\ref{fig:spectra} (top) shows the absorption coefficient of \ce{CO2} and other gases at standard temperature and pressure computed using the CIA and sublorentzian line broadening parametrization described in \cite{Wordsworth2010}, with the two regions of CIA indicated. In the lower panel, the outgoing longwave radiation (OLR) computed assuming a 1~bar dry \ce{CO2} atmosphere with surface temperature of 250~K is shown. As can be seen, the CIA causes significant reduction in OLR, but important window regions remain, particularly around 400 and 1000~cm$^{-1}$. The most effective minor greenhouse gases on early Mars are those whose absorption peaks in these \ce{CO2} window regions. Water vapour absorbs strongly at low wavenumbers and around its $\nu_2$ band at 1600~cm$^{-1}$, but its molar concentration is determined by temperature and hence it can only cause a feedback effect on the radiative forcing of other gases. At low temperature, this feedback is weak. Hence for pure clear-sky \ce{CO2}-\ce{H2O} atmospheres under early Martian conditions, modern radiative-convective models obtain mean surface temperatures of 225~K or less \citep{Wordsworth2010,Ramirez2014,Halevy2014}.

\begin{marginnote}
\entry{OLR}{Outgoing longwave radiation}
\end{marginnote}

\subsection{Alternative Long-term Warming Mechanisms}

The popular notion that Mars was once warm and wet combined with the impossibility of warming early Mars by \ce{CO2} alone has motivated investigation of many alternative warming mechanisms. For the most part, researchers have used one-dimensional radiative-convective models [either line-by-line or correlated-$k$; \cite{Goody1989}] to investigate the early climate. Radiative-convective models allow for temperature variations with altitude only and parametrize or neglect the effects of clouds, which strongly limits their accuracy and predictive power. Nonetheless, their speed and robustness make them invaluable tools for constraining parameter space and investigating novel effects.

\begin{marginnote}
\entry{line-by-line}{computationally costly approach to radiative transfer that resolves individual spectral lines}
\end{marginnote}

\begin{marginnote}
\entry{correlated-$k$}{computationally efficient approach to radiative transfer that calculates distributions of line intensity in large spectral bands, at some cost to accuracy}
\end{marginnote}

Over the years, researchers have investigated various greenhouse gas combinations to achieve a warm and wet early Martian climate \citep{Sagan1977,Postawko1986,Kasting1997,Haberle1998,von2013,Ramirez2014}. Methane might appear to be a promising Martian greenhouse gas given its strong radiative forcing on the present-day Earth. However, it is not an effective warming agent on early Mars because its first significant vibration-rotation band absorbs around 1300~cm$^{-1}$ --- a region too far from the peak of the Planck function in the 200-270~K temperature range to cause much change to outgoing longwave radiation (see Fig.~\ref{fig:spectra}). Gases such as ammonia and carbonyl sulphide have greater radiative potential but lack efficient formation mechanisms and are photochemically unstable in the Martian atmosphere, so they cannot have been present long-term in large quantities. 

Other researchers have looked at hydrogen as a greenhouse gas \citep{Sagan1977,Ramirez2014}.  \cite{Wordsworth2013c} showed that \ce{N2}-\ce{H2} CIA can cause significant warming on terrestrial planets even when \ce{H2} is a minor atmospheric constituent, because broadening of the CIA spectrum at moderate temperatures causes absorption to extend into window regions. In a thought-provoking paper, \cite{Ramirez2014} proposed that \ce{CO2}-\ce{H2} absorption could have caused warming on early Mars in a similar fashion, perhaps sufficiently to put the climate into a warm and wet state. Hydrogen readily escapes from a small planet like Mars, so to work this mechanism requires rapid hydrogen outgassing, which means a very reducing mantle and high rate of volcanism. It also requires a high atmospheric \ce{CO2} content, which as discussed in Section~\ref{subsec:howthick} may be inconsistent with a highly reducing mantle  \citep{HirschmannWithers2008}. Finally, a long-lived highly reducing atmosphere is not obviously consistent with evidence for oxidizing surface conditions during  the Noachian \citep{Chevrier2007}. Nonetheless, the contribution of reducing species to the early Martian climate and atmospheric chemistry is an interesting subject that requires further research.

\begin{figure}
\includegraphics[width=4.5in]{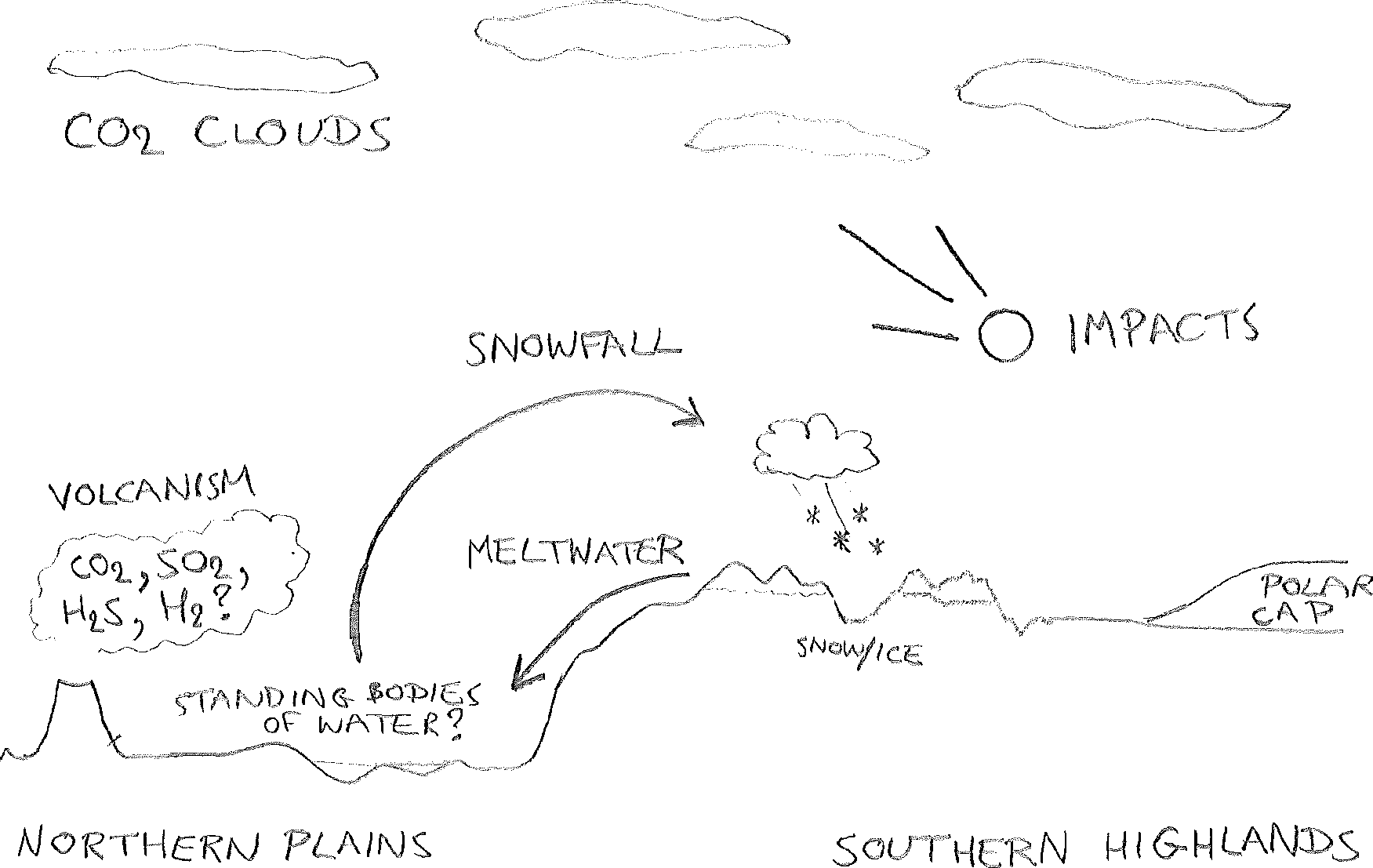}
\caption{Schematic of the major climate processes on Mars in the Noachian and early Hesperian periods. This cartoon assumes an episodically warm scenario for the early climate with long-term transport of snow to the southern highlands interrupted by episodic melting events.}\label{fig:schematic}
\end{figure}

The radiative forcing of clouds and aerosols was certainly also important to the early Martian climate, but it is challenging to constrain. One novel feature of cold \ce{CO2} atmospheres is that condensation at high altitudes leads to \ce{CO2} cloud formation (Fig.~\ref{fig:schematic}) --- an effect that is observed in the Martian mesosphere today \citep{Montmessin2007}. \cite{Forget1997} proposed that infrared scattering by \ce{CO2} clouds in the high atmosphere could have led to significant, long-term greenhouse warming on early Mars. However, to be effective this warming mechanism requires cloud coverage close to 100\%. Recent three-dimensional global climate modeling \citep{Forget2013,Wordsworth2013a} has shown that this level is never reached in practice. In addition, recent multiple-stream scattering studies have indicated that the two-stream methods used previously to calculate \ce{CO2} cloud climate effects tend to overestimate the strength of the scattering greenhouse \citep{Kitzmann2013}.  Hence the warming effect of \ce{CO2} clouds is likely to have been small in reality. Nonetheless, the infrared scattering effect is still an important term in their overall radiative budget. This means that they at least do not dramatically cool the climate via shortwave scattering, as was thought to be the case in the earliest studies of \ce{CO2} condensation on early Mars \citep{Kasting1991}. 

Recent studies have also investigated the role of \ce{H2O} clouds. In a 3D climate model study, \cite{Urata2013} found high-altitude water clouds formed that substantially decreased the OLR. They proposed that this could have caused transitory or long-lived warm climate states on early Mars. However, another 3D climate study published in the same year as Urata \& Toon's work found much less effective upwards transport of water vapor, resulting in mainly low-lying \ce{H2O} clouds that cooled the planet by increasing the albedo  \citep{Wordsworth2013a}. It is not unduly surprising that two 3D models of early Mars produce such different results on cloud forcing, given the uncertainty on this issue for the present-day Earth \citep{Forster2007}. Nonetheless, the discrepancy highlights the need for future detailed study of this issue.

\subsection{Episodic Warming}\label{subsec:episodic}

The geologic evidence that Mars once had large amounts of surface liquid water is conclusive, but  geomorphic constraints on the \emph{duration} for which that water flowed are much weaker. In addition, much of the geochemical evidence points towards surface conditions that were not warm and wet for long time periods. This is important, because if repeated transient melting events are capable of explaining the observations, the theoretical possibilities for warming become greater.

\subsubsection{Impact-induced Steam Atmospheres} The late Noachian is coincident with the early period of intense impactor flux known as the late heavy bombardment \citep[e.g.,][]{Hartmann2001}. Meteorite impacts were hence unquestionably a major feature of the environment during the main period of valley network formation. Based on this, several proposals for impact-induced warming have been put forward. \cite{Segura2002,Segura2008} suggested that large impactors could have evaporated up to tens of meters of water into the atmosphere, which would then have caused erosion when it rained back down to the surface. Later, \cite{Segura2012} proposed that impact-induced atmospheres could be very long-lived due to a strong decrease in OLR with surface temperature in a steam atmosphere, which would give rise to a climate bistability.

Despite the appealing temporal correlation, impact-induced steam atmospheres are not compelling as the main explanation for valley network formation. There are two main reasons for this. First, for transient impact-driven warming, there appears to be a large discrepancy between the estimated valley network erosion rates \citep{Barnhart2009,Hoke2011} and the amount of rainfall produced post-impact \citep{Toon2010}. Second, the runaway greenhouse bistability argument does not seem physically plausible, at least for clear-sky atmospheres, because it relies on the occurrence of extreme supersaturation of water vapor in the low atmosphere \citep{Nakajima1992}. 
If impacts played a role in carving the valley networks, therefore, they must have done so by a more indirect method.

\subsubsection{Sulfur-bearing Volcanic Gases}\label{subsec:sulf}

Another idea that has seen intensive study over the last few decades is the \ce{SO2}/\ce{H2S} greenhouse. The martian surface is sulfur-rich \citep{Clark1976} and sulphates are abundant on Hesperian and Noachian terrain \citep{Gendrin2005,Bibring2005,Ehlmann2014}. This suggests that volcanic emissions of sulfur-bearing gases (\ce{SO2} and \ce{H2S}) could have had a significant effect on early climate \citep{Postawko1986,Yung1997,Halevy2007}.

\ce{SO2} is a moderately effective greenhouse gas. The 518~cm$^{-1}$ (19.3~$\upmu$m) vibrational-rotation band associated with its $\nu_2$ bending mode absorbs close the peak of the blackbody spectrum at 250-300~K, but sufficiently far from the \ce{CO2} $\nu_2$ band at 667~cm$^{-1}$ to cause a fairly large reduction in the OLR if \ce{SO2} is present at levels of 10~ppm or above (Fig.~\ref{fig:spectra}). \ce{SO2} absorption bands in the 1000-1500~cm$^{-1}$ region also contribute but partially intersect with \ce{CO2} CIA at high pressure. \ce{H2S}, which would also have been outgassed in significant quantities by the reducing Martian mantle, is a far less effective greenhouse gas on early Mars due to the intersection of its main absorption bands with \ce{H2O} and \ce{CO2} (Fig.~\ref{fig:spectra}).

Like \ce{NH3} and \ce{CH4}, \ce{SO2} is photolyzed in the Martian atmosphere. Photochemical modeling under plausible early Martian conditions has suggested that this limits its lifetime to under a few hundred years \citep{Johnson2009}. More importantly, \ce{SO2} photochemistry leads to rapid formation of sulfate aerosols \citep{Tian2010}. These scatter incoming sunlight effectively, raising the albedo and cooling the planet. Similar climate effects have been observed in stratovolcano eruptions on Earth, such as that of Pinatubo in 1991 \citep{Stenchikov1998}.

\cite{Halevy2014} argued that intense episodic volcanic \ce{SO2} emissions associated with the formation of Hesperian ridged plains could have caused significant greenhouse warming. Using a line-by-line radiative-convective climate model, they found subsolar zonal average temperatures of 250~K for \ce{SO2} concentrations of 1-2~ppm in a clear-sky \ce{CO2}-\ce{H2O} atmosphere. Based on this, they argued that peak daytime equatorial temperatures would exceed 273~K for several months a year during transient pulses of volcanism and hence significant meltwater could be generated. 

Because \cite{Halevy2014} used a one-dimensional column model, they did not account for horizontal heat transport by the atmosphere. In contrast to their result, recent 3D GCM studies find that in concentrations of under 10~ppm, \ce{SO2} warming cannot cause significant melting events on early Mars \citep{Mischna2013,Kerber2015}. Indeed, the dramatic cooling effects of sulfate aerosols together with the timing of the Hesperian flood basalts led \cite{Kerber2015} to suggest an opposite conclusion: the onset of intense sulfur outgassing on Mars may have \emph{ended} the period of episodically or continuously  warm conditions in the late Noachian.

In summary, no single mechanism is currently accepted as the cause of anomalous warming events on early Mars. Climate models that allow horizontal temperature variations  show that in combination, various atmospheric and orbital effects can combine to create marginally warm conditions and hence small amounts of episodic melting \citep{Richardson2005,Wordsworth2013a,Kite2013,Mischna2013,Wordsworth2015}, particularly if the meltwater is assumed to be briny \citep{Fairen2009,Fairen2010}. Just like the climate of Earth today, the ancient climate of Mars was probably complex, with multiple factors contributing to the mean surface temperature. Nonetheless, further research on this key issue is necessary. The continuing uncertainty with regard to warming mechanisms has recently led some studies to take an empirical approach to constraining the early Martian climate (Section~\ref{subsec:warmwetmodel}). 

\section{DECIPHERING THE LATE NOACHIAN WATER CYCLE}\label{sec:watercycle}

Radiative-convective climate models are powerful tools, but they have limitations. 
One of the most obvious is their inability to capture cloud effects, except in a crude and highly parameterized way. A second major limitation is that they fail to account for regional differences in climate, and hence cannot be used to model a planet's hydrological cycle. This is particularly important for Mars, which has large topographic variations and a spatially inhomogenous surface record of alteration by liquid water.

\begin{marginnote}
\entry{GCM}{general circulation model}
\end{marginnote}

While 3D general circulation models (GCMs) of the present-day Martian atmosphere began to be used from the 1960s onwards \citep{Leovy1969}, development of GCMs for paleoclimate applications has been much slower. An important reason for this is the complexity of dense gas \ce{CO2} radiative transfer, as described above. Nonetheless, in the last five years, several teams have begun 3D GCM modeling of the early Martian climate.

Challenges in simulating the Martian paleoclimate in three dimensions include a potentially altered topography compared to present-day, uncertainty in the orbital eccentricity and obliquity, and the difficulty of calculating radiative transfer accurately and rapidly in an atmosphere of poorly constrained composition. Of all of these, the latter poses the greatest technical challenge. Line-by-line codes such as that used to produce Fig.~\ref{fig:spectra} are impractical for GCM simulations because the number of calculations required makes them prohibitively expensive computationally. Instead, recent 3D simulations have used the correlated-$k$ method \citep{Goody1989,Lacis1991}. This technique, which was originally developed for terrestrial radiative transfer applications, replaces the line-by-line integration over spectral wavenumber with a sum over a cumulative probability distribution in a much smaller number of bands.

The first study to apply this technique to 3D Martian paleoclimate simulations was \cite{StewartJ2008}, who looked at the effect of \ce{SO2} warming from volcanism. While pioneering in terms of its technical approach, this work used correlated-$k$ coefficients that were later found to be in error \citep{Mischna2013}. As a result, it predicted unrealistically high warming due to \ce{SO2} emissions. Since this time, several new 3D GCM studies of early Mars using correlated-$k$ radiative transfer have been published, leading to a number of new insights. 

As previously discussed, \cite{Forget2013}, \cite{Wordsworth2013a} and \cite{Urata2013} investigated the role of \ce{CO2} and \ce{H2O} clouds in warming the early climate, and found both cloud fraction and mean particle size were critical factors in their radiative effect.  \cite{Forget2013} and \cite{Soto2015} investigated collapse of \ce{CO2} atmospheres due to surface condensation. They found that the process was extremely significant at low obliquities (below 20$^\circ$) and at pressures above 3~bar. The predictions of \cite{Forget2013} and \cite{Soto2015} differed in detail, however, partly because \cite{Soto2015} neglected the effects of \ce{CO2} clouds and used the present-day solar luminosity. Hence further model intercomparison on this issue is required. In a related study, \cite{Kahre2013} examined \ce{CO2} collapse in the presence of an active dust cycle and found that dust could help to stabilize moderately dense atmospheres ($\sim 80$~mbar) against collapse at high obliquity. As described in Section~\ref{subsec:sulf}, two new 3D studies have also investigated the role of \ce{SO2} warming in the Noachian climate \citep{Mischna2013,Kerber2015}.

\subsection{Adiabatic Cooling and the Icy Highlands Hypothesis}

Another outcome of recent 3D GCM modeling has been an increased understanding of the processes governing Mars' early surface water cycle. On a cold planet, the water cycle is dominated by the transport of surface ice to regions of enhanced stability (cold traps). Ice stability is a strong function of the sublimation rate, which depends exponentially on surface temperature via the Clausius-Clayperon relation. In practice, this means that the regions of the planet with the lowest annual mean surface temperatures are usually cold traps. Figuring out the cold trap locations on early Mars is critical, because it tells us where the water sources were during warming episodes.

Mars today has an atmospheric pressure of around 600~Pa and obliquity of 25.2$^\circ$. The majority of surface and subsurface water ice is found near the poles \citep{Boynton2002}. Mars' obliquity has varied significantly throughout the Amazonian, however \citep{Laskar1993}, and at high obliquities ice migration to equatorial \citep{Mischna2003,Forget2006} and mid-latitude \citep{Madeleine2009} regions is predicted. Obliquity variations may well have also been important to ice stability in the Noachian and early Hesperian. However, in this period the role of atmospheric pressure was probably even more important.

\begin{figure}
\includegraphics[width=4.5in]{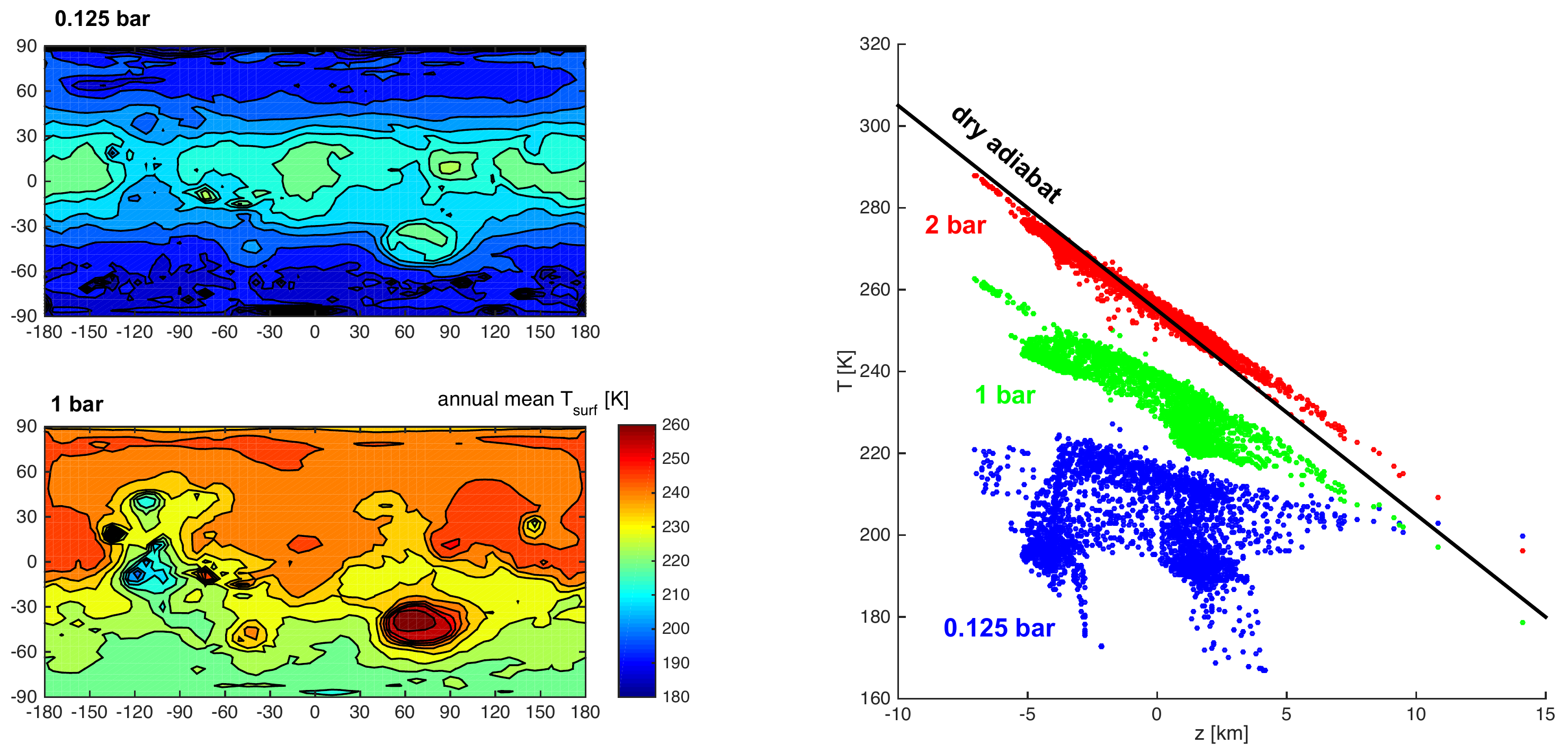}
\caption{The adiabatic cooling effect on early Mars. (top left) Annual mean surface temperature from a 3D GCM simulation with 0.125~bar surface pressure. (bottom left) Annual mean surface temperature from a 3D GCM simulation with 1~bar surface pressure. (right) Scatter plot of surface temperature vs. altitude for simulations with 0.125, 1 and 2 bar surface pressure. The dry adiabat $g/c_p$ is also indicated. Data for the plots was acquired from the 41.8$^\circ$ obliquity, fixed relative humidity simulations described in \cite{Wordsworth2015} [see also \cite{Forget2013}].
}\label{fig:adia_cool}
\end{figure}

Figure~\ref{fig:adia_cool} shows the simulated annual mean temperature $T_s$ for Mars given a solar flux 75\% of that today \citep{Gough1981} and surface pressures of a) $p_s = 0.125$~bar and $p_s = 1$~bar. At the lowest pressure, mean surface temperatures are primarily determined by insolation and  the main variation of $T_s$ is with latitude. At 1~bar, however, a  shift to variation of $T_s$ with altitude has occurred. The accompanying scatter plot of surface temperature vs. altitude clearly shows a trend towards temperature-altitude anticorrelation as pressure increases.
The origin of this effect is the increase in coupling between the lower atmosphere and surface via the planetary boundary layer.

The surface heat budget on a mainly dry planet can be written as 
\begin{equation}
F_{lw,\uparrow} = F_{lw,\downarrow} + F_{sw,\downarrow} + F_{sens} 
\end{equation}
where $F_{lw,\uparrow}$ is the upwelling longwave (thermal) radiation from the surface, $F_{lw,\downarrow}$ is the downwelling thermal radiation from the atmosphere to the surface, $F_{sw,\downarrow}$ is the incoming solar flux, and $F_{sens}$ is the sensible heat exchange. The latter term can be written as\footnote{At low atmospheric pressures ($<0.2$~bar) and surface wind speeds free convection may also be important [\cite{Kite2013}; Fig.~9], but its relative effect declines at higher pressures.} $F_{sens}=\rho_a C_D c_p |\mathbf u |(T_a-T_s)$, where $\rho_a$ is the atmospheric density near the surface, $C_D$ is the bulk drag coefficient, $c_p$ is the atmospheric specific heat capacity at constant pressure, $|\mathbf u |$  is the surface windspeed and $T_a$ is the temperature of the atmosphere at the surface. Observations and simulations indicate that $|\mathbf u |$ generally decreases with $\rho_a$, but only slowly. Hence the magnitude of $F_{sens}$ will increase significantly with $\rho_a$ unless the temperature difference $T_a-T_s$ simultaneously decreases.

For a planet without an atmosphere (such as Mercury), $F_{lw,\downarrow}$ and $F_{sens}$ equal zero and surface temperature is determined by solar insolation (with a contribution from geothermal heating in very cold regions). As atmospheric pressure increases, so does heat exchange between the atmosphere and surface. For a planet with a thick atmosphere (such as Venus or Earth), sensible and radiative atmospheric heat exchange are significant and drive $T_s$ towards the local air temperature $T_a$. Because the atmospheric lapse rate follows a convective adiabat in the troposphere, surface temperatures therefore decrease with altitude. This is exactly the effect that causes equatorial mountains on Earth such as Kilimanjaro to have snowy peaks despite the tropical temperatures at their bases.

Mars has a lower surface gravity than Earth, which decreases its dry adiabatic lapse rate [$\Gamma_d = -g/c_p$, with $g$ gravity and $c_p$ the specific heat capacity at constant volume]. However, this is more than compensated for by its large topographic variations. The altitude difference between Hellas Basin and the southern equatorial highlands is around 12~km, or just over one atmospheric scale height. In 3D GCM simulations, the corresponding drop in annual mean temperature is around 30~K at 1~bar atmospheric pressure, or around 40~K at 2~bar (Fig.~\ref{fig:adia_cool}).

As a result of this adiabatic cooling effect, for moderate values of Martian obliquity and atmospheric pressures above around 0.5~bar, the equatorial Noachian highlands (where most valley networks are observed) become cold traps, confirming a prior prediction by \cite{Fastook2012}. 
Long-term three-dimensional climate simulations coupled to a simple ice evolution model \citep{Wordsworth2013a} have demonstrated that as a result, ice migrates to the valley network source regions regardless of where it is initially located on the surface.

The adiabatic cooling effect led Wordsworth et al. to propose the so-called ``icy highlands'' scenario for the early climate (Fig.~\ref{fig:schematic}). In essence, the idea is that if the valley network source regions were cold traps, the early Martian water cycle could have behaved somewhat like a transiently forced, overdamped oscillator. Episodic melting events (the perturbing force), would have transported  \ce{H2O} to lower altitude regions of the planet on relatively short timescales. Over longer timescales, adiabatic cooling (the restoring mechanism) would have returned the system to equilibrium.

Recent modeling and observational work has used the icy highlands hypothesis as a framework for testing various predictions on the early Martian climate. For example, \cite{Scanlon2013} used an analytical model combined with the 3D GCM of \cite{Wordsworth2013a} to study orographic precipitation over Warrego Vallis and were able to match local precipitation patterns with the valley network locations. \cite{Head2014} discussed the analogies between the icy highlands scenario for early Mars and the Antarctic Dry Valleys, which have long been considered one of the most ``Mars-like'' regions on Earth.

 \begin{marginnote}
\entry{firn}{a partially compacted form of solid \ce{H2O} intermediate in density between snow and ice.}
\end{marginnote}

 \cite{Fastook2015} used a glacial flow model to study how the buildup of large ice sheets would have affected the geomorphology of the Noachian highlands. Assuming a thermal conductivity appropriate to ice without a blanketing effect from snow or firn, they found that if the Noachian water inventory was $< 5 \times $ the present-day GEL (i.e. 170~m),  equatorial highland glaciers would have been cold-based and hence would not have left traces on the surface in the form of cirques, eskers or other glacial landforms. This conclusion was broadly confirmed by \cite{Cassanelli2015b}, who studied the influence of snow and firn thermal blanketing on ice sheet melting under a \ce{H2O}-limited scenario only. The relative lack of glaciation in the equatorial highlands and the total water inventory are key pieces of the Noachian climate puzzle that we return to in Section~\ref{subsec:peripara}.

\subsection{The Hydrological Cycle on a Warm and Wet Planet}\label{subsec:warmwetmodel}

The icy highlands scenario provides a useful working model for thinking about the early Martian climate and fits many aspects of the geologic evidence. Nonetheless, because some observations are still interpreted as supporting long-term warm and wet conditions, it is also interesting to model alternative scenarios for the early Martian climate. As we have discussed, no climate model based on realistic assumptions currently predicts warm and wet conditions for early Mars\footnote{\cite{Ramirez2014} argue that the \ce{CO2}-\ce{H2} CIA mechanism can come close. However, in their model it still requires more atmospheric \ce{H2} than they can produce, even using their most generous outgassing estimates.}. To study the hydrological cycle on a warm and wet planet, therefore, it is necessary to use an empirical approach. There are two basic ways to do this: increasing the solar luminosity, or increasing the atmospheric infrared opacity.

\cite{Wordsworth2015} tried both these approaches to study precipitation patterns on a warm and wet early Mars. As a starting condition, they assumed that Mars had a global northern ocean and smaller bodies of water in Argyre and Hellas, based on the putative delta shoreline constraints of \cite{diAchille2010}. For simplicity, they also neglected the destabilizing effects of the ice-albedo feedback on the early Martian climate (Section~\ref{subsec:snowball}). 

In general, \cite{Wordsworth2015} found that the precipitation patterns in the warm and wet case were not a close match to the valley network distribution. In particular, the presence of Tharsis caused a dynamical effect on the circulation that led to low rainfall rates in Margaritifer Sinus, where some of the most well-developed valley networks on Mars are found. This indicates that either a) something was missing from their model or b) Mars was never warm and wet and the Martian valley networks formed through transient melting events. The opinion of this author is that the second possibility is the correct one. Future work testing the influence of poorly constrained effects such as cloud convection parametrizations, changes in surface topography and possibly true polar wander on this result will allow the first possibility to be tested further. In any case, it is clear that systematic empirical investigation of different scenarios using 3D models provides a new way to constrain the early climate.

\subsection{Snowball Mars}\label{subsec:snowball}

An additional impediment to the warm and wet scenario for early Mars that has not yet been extensively considered is the ice-albedo feedback \citep{Budyko1969}. This process is an important player in the ongoing loss of sea ice in the Arctic on Earth due to anthropogenic climate change \citep{Stroeve2007}. It was also responsible for the snowball Earth global glaciations that occurred in the Neoproterozoic \citep{Kirschvink1992,Hoffman1998,Pierrehumbert2011_NP}.

Even with a northern ocean at the  \cite{diAchille2010} shoreline, Mars would possess a far higher land to ocean ratio than Earth does, making the physics of a snowball transition quite different. Unlike on Earth, where runaway glaciation occurs through freezing of a near-global ocean, on Mars transport of \ce{H2O} to high altitude regions as snow would play the key role. 
Because of the presence of Tharsis at the equator, the topography of Mars is particularly conducive to an ice-albedo instability of this kind. As discussed in Section~\ref{sec:geo_evid}, most of the formation of Tharsis was probably complete by the late Noachian (see also Fig.~\ref{fig:timeline}). Thanks to the adiabatic cooling effect under a thicker atmosphere, Tharsis is an effective cold trap for water ice even if most of Mars is assumed to be warm and wet. 

To put numbers to this idea, we can define sea level as -2.54~km from the datum following \cite{diAchille2010} and take the summit of Tharsis (neglecting the peaks of the Tharsis Montes volcanoes, which are Amazonian-era \citep{Tanaka2014}) to be approximately $z=8$~km. Then the sea-level-to-summit adiabatic temperature difference is $\Gamma_d z = -g z/c_p \approx 38$~K. At 1~bar surface pressure the surface temperature gradient with altitude is still somewhat below the adiabatic value, so we can conservatively estimate a temperature difference of 30~K. To a first approximation, seasonal temperature changes at the equator can be neglected, so an annual mean sea level temperature of around 30$^\circ$C is required to avoid the buildup of snow and ice on Tharsis. 

The high altitude of Tharsis means that the atmospheric radiative effects above it (both scattering and absorption) are reduced. This increases its radiative forcing due to surface albedo changes compared to lower-lying regions. Because of its equatorial location, an ice-covered Tharsis also increases the planetary albedo regardless of Mars' obliquity. 

No 3D climate simulation has yet addressed the impact of a snowy Tharsis on the stability of a warm and wet early Mars. This is an important topic for future study. 
It may be that marginally warm global mean temperatures are insufficient to stop Mars from rapidly transitioning to a cold and wet state.

\subsection{The periglacial paradox and the Noachian surface water inventory}\label{subsec:peripara}

As discussed in Section~\ref{sec:geo_evid}, there is little evidence for glaciation during Mars' Noachian period in the equatorial highlands. At first glance, this seems like a potential drawback of  the icy highlands scenario for the early climate. If snow and ice buildup in high altitude regions was the source of the water that carved the valley networks, should we not see a more widespread record of glacial and fluvioglacial alteration across the Noachian highlands?

The trouble with this line of thinking as an argument against the icy highlands scenario is that it equally applies to a warm and wet early Mars. If Mars was once warm and wet, it must have subsequently become cold, because it is cold today. Once it did, liquid water would freeze into ice and migrate to the cold trap regions of the surface. This implies the buildup of ice sheets in the equatorial highlands unless the Martian atmospheric pressure immediately decreased to low values \emph{and} obliquity was continually low in the period immediately following the warm and wet phase (which seems unlikely). The quantity of surface water sufficient to fill a northern ocean to the \cite{diAchille2010} shoreline implies a GEL of around 550~m, so in the immediate post-warm-and-wet phase these ice sheets could be several kilometers thick.  As shown by \cite{Fastook2015}, this would lead to wet-based glaciation (and hence fluvioglacial erosion) even under very climate cold conditions, and even without the insulating effects of a surface snow layer.

If a warm and wet Mars should leave abundant evidence of glaciation in its subsequent cold and wet phase, how can the equatorial periglacial paradox be explained?  Most likely,  the resolution lies in early Mars' total surface \ce{H2O} inventory. In a supply-limited icy highlands scenario with episodic melting, snow and ice deposits are cold-based. Then, the only significant alteration of terrain comes from fluvial erosion during melting events. 

Many studies have investigated the evolution of Mars' surface \ce{H2O} inventory through time. The deuterium to hydrogen (D/H) ratio can be used to constrain the early Martian water inventory \citep{Greenwood2008,Webster2013,Villanueva2015}, although our lack of knowledge of the dominant escape process in the Noachian/Hesperian and of the cometary contribution to Mars' surface water \citep{Marty2012} complicates the analysis. \cite{Villanueva2015} recently used Earth-based observations of Martian D/H and estimated the late Noachian water GEL to be 137~m. In contrast, \cite{Carr2015} recently calculated the \ce{H2O} loss/gain budget in the Hesperian and Amazonian and concluded that the late Noachian water GEL was as low as 24~m. Although the uncertainty is considerable, most estimates place the early Martian water inventory below a few hundred meters GEL. This low inventory compared to Earth's is consistent with Mars' low mass and likely significant loss of atmosphere to space (see sidebar).

\begin{figure}
\includegraphics[width=4.5in]{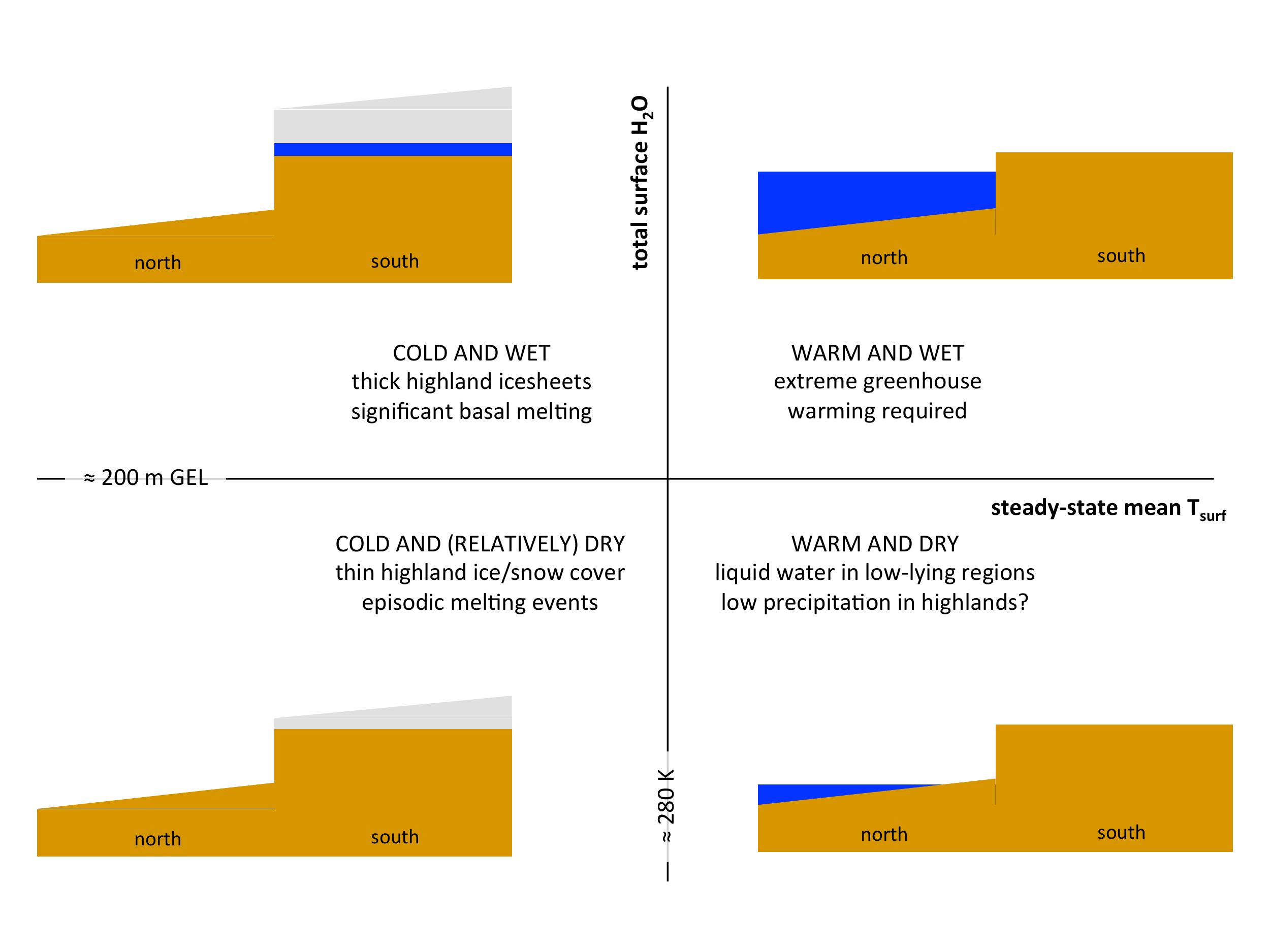}
\caption{({\it Left column}) Conceptual phase diagram for the \emph{steady-state} early Martian climate under a denser atmosphere. Given mean surface temperature and the total surface \ce{H2O} inventory as $x$ and $y$ axes, respectively, four idealized climate states can be imagined. The cold and wet state leads to large highland wet-based glacial icesheets, in conflict with the geologic record. The warm and wet state requires extreme greenhouse warming and rapidly transitions to the cold and wet state when surface temperatures decrease. The warm and dry state has not yet been simulated numerically but may lead to little or no rainfall in the equatorial highlands. The cold and relatively dry state, when combined with episodic melting events, appears consistent with the majority of the geologic evidence.
}\label{fig:phase_diag}
\end{figure}

Continuing our previous line of thought to its logical conclusion, we can construct an idealized two dimensional phase diagram for the early Martian climate, with surface temperature on one axis and the total surface \ce{H2O} inventory on the other (Figure~\ref{fig:phase_diag}). Each quadrant in Fig.~\ref{fig:phase_diag} represents a different  end-member state for the long-term climate and surface hydrology of early Mars. The warm and wet state with a northern ocean is disfavored for the geochemical and climatological reasons discussed previously. Because of the ice-albedo feedback, it also readily transitions to a cold and wet state. The cold and wet state implies extensive wet-based glaciation across Noachian terrain, in conflict with the geomorphological record. The cold and (relatively) dry state (water GEL$<$200~m) is essentially the icy highlands scenario, which in combination with the right amount of episodic melting can explain most of the geologic record. Finally, there is a possible ``warm and dry'' scenario where all the surface water is liquid but the \ce{H2O} GEL is below $\sim$200~m is also possible. Potentially, this might also fit many of the geologic constraints on the early Martian climate. However, it is climatologically as hard to justify as the warm and wet state. In addition, in such a scenario the low-lying regions where liquid water stabilizes would be so far from the valley network source regions that precipitation there might be limited or non-existent. Preliminary GCM studies using the model described in \cite{Wordsworth2015} suggest that this is indeed the case. This is yet another important issue that deserves to be studied in detail in future.

\begin{textbox}
\section{MARS, THE RUNT PLANET}
Mars is fascinating to study in its own right, but also because because of the insight it can give us into planetary evolution and habitability in general. Several potentially rocky exoplanets and exoplanet candidates that receive approximately the same stellar flux as Mars have now been discovered [e.g., \cite{Udry2007,Quintana2014}]. Can we derive lessons from Martian climate evolution studies that are generalizable to a wider context?

Mars formed very rapidly compared to Earth \citep{Dauphas2011} and is smaller than predicted by standard formation models \citep{Chambers1998}. Current thinking suggests Mars is best understood as a planetary embryo, whose development into a fully fledged terrestrial planet was arrested by some process such as instability in the configuration of the outer planets \citep{Walsh2011}. The Red Planet is not as massive as it could have been, and this factor more than any other has dominated its subsequent evolution.

Mars' low mass relative to Earth led to an early shutdown of the magnetic dynamo \citep{Acuna1998} and rapid cooling of the interior. If plate tectonics ever initiated at all, it ceased rapidly \citep{Connerney2001,Solomon2005}. As a result, volatile cycling between the surface and interior was  strongly inhibited and the rate of atmospheric loss to space was probably also enhanced. The present-day atmosphere is so thin that liquid water is unstable on the surface. Mars' reddish, hyperoxidised surface, which is extremely inhospitable to life, is a direct result of the escape of hydrogen to space over geologic time \citep{Lammer2003}. 

Almost certainly, Mars tells us more about the habitability of low mass planets than the habitability of planets that are far from their host stars. Indeed, an Earth-mass planet with plate tectonics at Mars' orbital distance would be habitable today, if mainstream thinking on the carbonate-silicate cycle  \citep{Walker1981} and planetary habitability \citep{Kasting1993} is correct. Radiative-convective calculations using the model described in \cite{Wordsworth2013c} indicate that a 1~$M_\oplus$ planet at Mars' orbit with an atmosphere dominated by \ce{CO2} and \ce{H2O} would have a surface temperature of 288~K for an atmospheric pressure of around 3-5~bar -- a small amount in comparison with Earth's total carbon inventory \citep{Sleep2001,Hayes2006}. 

In the absence of the still-mysterious process that abruptly halted Mars' growth, our more distant neighbour could have been globally habitable to microbial life through much of its history. Given the potential for exchange of biological material on impact ejecta between terrestrial planets \citep{Mileikowsky2000}, biogenesis on Earth [or Mars; \cite{Kirschvink2002}] would then have led to the development of global biospheres on two planets in the Solar System. A future scenario in which Mars possesses a global biosphere is also possible, but will depend on human colonization and subsequent intentional modification of the climate.
\end{textbox}

\section{OUTLOOK}

While major questions remain on the nature of the early Martian climate, recent advances have been significant. The weight of geomorphological and geochemical evidence points towards a late Noachian hydrological cycle that was intermittent, not permanently active. Three-dimensional GCM simulations of the early climate and other modeling and analogue studies suggest a water-limited early Mars with episodic melting episodes may be a suitable paradigm for much of the late Noachian and early Hesperian climate. 

Despite the progress over the last few years, aspects of the early climate remain unclear. Chief among these is the driving mechanism behind the episodic warming events. Previous theories for warming by \ce{CO2} clouds, sulfur volcanism and impact-induced steam atmospheres have all been shown to possess serious problems. Indeed, as one- and three-dimensional climate models become more accurate, there has been a general trend towards prediction of colder surface temperatures. A successful theory for surface warming must yield valley network erosion rates consistent with the geologic evidence \citep{Hoke2011} but avoid extensive surface carbonate formation and aqueous alteration of sediments   \citep{Tosca2009,Ehlmann2011}.

The climate of early Mars, like that of present-day Earth, almost certainly involved multiple interacting processes. In addition, despite the difficulties with the impact-induced steam atmosphere hypothesis, the timing of the peak period of valley network formation with respect to the late heavy bombardment  is suggestive of a causal link. In contrast, the lack of temporal correlation between the valley networks and Hesperian ridged plains argues against a causal link with effusive volcanism.

The key advantage of modeling the early climate as a spatially varying system, rather than studying isolated processes or trying to achieve mean temperatures above 273~K in a globally averaged model is that it allows tighter intercomparison with the geologic evidence. Future research should work towards close integration of surface geology, 3D climate and hydrological modeling studies. Specific regions where this approach would be particularly useful include the south pole around the Dorsa Argentea Formation and the Aeolis quadrangle where Gale crater is located. Issues of specific importance to the 3D climate modeling include better representation of clouds, perhaps through a sub-gridscale parametrization scheme [e.g., \cite{Khairoutdinov2001}] and coupling with subsurface hydrology models \citep{Clifford1993,Clifford2001}.

If Mars never had a steady-state warm and wet climate, does this spell doom for the search for past life on the surface? Probably not: life on Earth clings tenaciously to almost any environment where we can look for it, including Antarctica's Dry Valleys and deep below the seafloor \citep{Cary2010,dHondt2004}. Indeed, if one view of Earth's climate in the Hadean and early Archean is correct, our own planet may have been in a globally glaciated state when life first formed \citep{Sleep2001}. The search for life on Mars must continue, but to maximize the chances of success it needs to be informed by our evolving understanding of the early climate.

\begin{marginnote}
\entry{MOLA}{Mars Orbiter Laser Altimeter}
\end{marginnote}

\begin{summary}[SUMMARY POINTS]
\begin{enumerate}
\item Mars underwent an extended period of surface erosion and chemical weathering by liquid water until around 3.5~Ga, during the late Noachian and early Hesperian periods.
\item The weight of the observational evidence favors a mainly cold climate with episodic warming events, rather than permanently warm and wet conditions.
\item If early Mars was once warm and wet, thick wet-based icesheets would have formed on the Noachian highlands when the warm period ended, causing significant glacial erosion.
\item Constraints on the early solar luminosity, Martian orbit and radiative transfer of \ce{CO2} strongly disfavor a warmer climate due to \ce{CO2} and \ce{H2O} only.
\item Under a thicker atmosphere, adiabatic cooling of the surface causes transport of snow and ice to the valley network source regions.
\item Repeated episodic warming events probably caused ice and snowpacks in the Noachian highlands to melt, carving valley networks and other fluvial features.
\item The precise mechanism that caused the warming events is still poorly constrained. The two most likely forcing mechanisms are meteorite impacts and volcanism, although the details remain unclear.
\end{enumerate}
\end{summary}

\section*{DISCLOSURE STATEMENT}
The author is not aware of any affiliations, memberships, funding, or financial holdings that might be perceived as affecting the objectivity of this review.

\section*{ACKNOWLEDGMENTS}
The author thanks Raymond Pierrehumbert for a constructive review of this manuscript, and numerous colleagues for their critical feedback and advice on key aspects of the observations, including Bethany Ehlmann, Caleb Fassett, Jim Head, Francois Forget and Laura Kerber. Bob Haberle is also acknowledged for enlightening discussion of the ``warm and dry'' state for the early climate.


\begin{thebibliography}{}
\expandafter\ifx\csname natexlab\endcsname\relax\def\natexlab#1{#1}\fi

\bibitem[{Acuna et~al.(1998)Acuna, Connerney, Wasilewski, Lin, Anderson
  et~al.}]{Acuna1998}
Acuna M, Connerney J, Wasilewski P, Lin R, Anderson K, et~al. 1998.
Magnetic field and plasma observations at mars: Initial results of the mars
  global surveyor mission.
\textit{Science} 279:1676--1680

\bibitem[{{Baranov}, {Lafferty} \& {Fraser}(2004)}]{Baranov2004}
{Baranov} YI, {Lafferty} WJ, {Fraser} GT. 2004.
{Infrared spectrum of the continuum and dimer absorption in the vicinity of the
  O2 vibrational fundamental in O2/CO2 mixtures}.
\textit{J. Mol. Spectrosc.} 228:432--440

\bibitem[{{Barnhart}, {Howard} \& {Moore}(2009)}]{Barnhart2009}
{Barnhart} CJ, {Howard} AD, {Moore} JM. 2009.
{Long-term precipitation and late-stage valley network formation: Landform
  simulations of Parana Basin, Mars}.
\textit{Journal of Geophysical Research (Planets)} 114:E01003

\bibitem[{{Bibring} et~al.(2006){Bibring}, {Langevin}, {Mustard}, {Poulet},
  {Arvidson} et~al.}]{Bibring2006}
{Bibring} J, {Langevin} Y, {Mustard} JF, {Poulet} F, {Arvidson} R, et~al. 2006.
{Global Mineralogical and Aqueous Mars History Derived from OMEGA/Mars Express
  Data}.
\textit{Science} 312:400--404

\bibitem[{{Bibring} et~al.(2005){Bibring}, {Langevin}, {Gendrin}, {Gondet},
  {Poulet} et~al.}]{Bibring2005}
{Bibring} JP, {Langevin} Y, {Gendrin} A, {Gondet} B, {Poulet} F, et~al. 2005.
{Mars Surface Diversity as Revealed by the OMEGA/Mars Express Observations}.
\textit{Science} 307:1576--1581

\bibitem[{Boynton et~al.(2002)Boynton, Feldman, Squyres, Prettyman,
  Br{\"u}ckner et~al.}]{Boynton2002}
Boynton W, Feldman W, Squyres S, Prettyman T, Br{\"u}ckner J, et~al. 2002.
Distribution of hydrogen in the near surface of mars: Evidence for subsurface
  ice deposits.
\textit{science} 297:81--85

\bibitem[{Brady \& G{\'\i}slason(1997)}]{Brady1997}
Brady PV, G{\'\i}slason SR. 1997.
Seafloor weathering controls on atmospheric co 2 and global climate.
\textit{Geochimica et Cosmochimica Acta} 61:965--973

\bibitem[{{Budyko}(1969)}]{Budyko1969}
{Budyko} MI. 1969.
{The effect of solar radiation variations on the climate of the earth.}
\textit{Tellus} 21:611--619

\bibitem[{Bullock \& Moore(2007)}]{Bullock2007}
Bullock MA, Moore JM. 2007.
Atmospheric conditions on early mars and the missing layered carbonates.
\textit{Geophysical Research Letters} 34

\bibitem[{Cabrol \& Grin(1999)}]{Cabrol1999}
Cabrol NA, Grin EA. 1999.
Distribution, classification, and ages of martian impact crater lakes.
\textit{Icarus} 142:160--172

\bibitem[{Carr \& Head(2015)}]{Carr2015}
Carr M, Head J. 2015.
Martian surface/near-surface water inventory: Sources, sinks, and changes with
  time.
\textit{Geophysical Research Letters} 42:726--732

\bibitem[{Carr(1996)}]{Carr1996}
Carr MH. 1996.
Water on mars.
\textit{New York: Oxford University Press,| c1996} 1

\bibitem[{{Carr} \& {Head}(2003)}]{Carr2003}
{Carr} MH, {Head} JW. 2003.
{Basal melting of snow on early Mars: A possible origin of some valley
  networks}.
\textit{Geophysical Research Letters} 30:240000--1

\bibitem[{Carr \& Head(2010)}]{Carr2010}
Carr MH, Head JW. 2010.
Geologic history of mars.
\textit{Earth and Planetary Science Letters} 294:185--203

\bibitem[{Carter et~al.(2015)Carter, Loizeau, Mangold, Poulet \&
  Bibring}]{Carter2015}
Carter J, Loizeau D, Mangold N, Poulet F, Bibring JP. 2015.
Widespread surface weathering on early mars: A case for a warmer and wetter
  climate.
\textit{Icarus} 248:373--382

\bibitem[{{Carter} et~al.(2010){Carter}, {Poulet}, {Bibring} \&
  {Murchie}}]{Carter2010}
{Carter} J, {Poulet} F, {Bibring} J, {Murchie} S. 2010.
{Detection of Hydrated Silicates in Crustal Outcrops in the Northern Plains of
  Mars}.
\textit{Science} 328:1682--

\bibitem[{Carter et~al.(2013)Carter, Poulet, Bibring, Mangold \&
  Murchie}]{Carter2013}
Carter J, Poulet F, Bibring JP, Mangold N, Murchie S. 2013.
Hydrous minerals on mars as seen by the crism and omega imaging spectrometers:
  Updated global view.
\textit{Journal of Geophysical Research: Planets} 118:831--858

\bibitem[{{Cary} et~al.(2010){Cary}, {McDonald}, {Barrett} \&
  {Cowan}}]{Cary2010}
{Cary} SC, {McDonald} IR, {Barrett} JE, {Cowan} DA. 2010.
{On the rocks: the microbiology of Antarctic Dry Valley soils}.
\textit{Nature Reviews Microbiology} 8:129--138

\bibitem[{Cassanelli \& Head(2015)}]{Cassanelli2015b}
Cassanelli JP, Head JW. 2015.
Firn densification in a late noachian ``icy highlands'' mars: Implications for
  ice sheet evolution and thermal response.
\textit{Icarus} 253:243--255

\bibitem[{Chambers \& Wetherill(1998)}]{Chambers1998}
Chambers JE, Wetherill GW. 1998.
Making the terrestrial planets: N-body integrations of planetary embryos in
  three dimensions.
\textit{Icarus} 136:304--327

\bibitem[{{Chassefi{\`e}re} \& {Leblanc}(2004)}]{Chassefiere2004}
{Chassefi{\`e}re} E, {Leblanc} F. 2004.
{Mars atmospheric escape and evolution; interaction with the solar wind}.
\textit{Planetary and Space Science} 52:1039--1058

\bibitem[{Chevrier, Poulet \& Bibring(2007)}]{Chevrier2007}
Chevrier V, Poulet F, Bibring JP. 2007.
Early geochemical environment of mars as determined from thermodynamics of
  phyllosilicates.
\textit{Nature} 448:60--63

\bibitem[{{Clark} et~al.(1976){Clark}, Baird, {Rose}, Toulmin, Keil
  et~al.}]{Clark1976}
{Clark} BC, Baird AK, {Rose} HJ, Toulmin P, Keil K, et~al. 1976.
Inorganic analyses of martian surface samples at the viking landing sites.
\textit{Science} 194:1283--1288

\bibitem[{{Clifford}(1993)}]{Clifford1993}
{Clifford} SM. 1993.
A model for the hydrologic and climatic behavior of water on mars.
\textit{Journal of Geophysical Research} 98:10--973

\bibitem[{{Clifford} \& {Parker}(2001)}]{Clifford2001}
{Clifford} SM, {Parker} TJ. 2001.
{The evolution of the martian hydrosphere: Implications for the fate of a
  primordial ocean and the current state of the northern plains}.
\textit{Icarus} 154:40--79

\bibitem[{Clough, Iacono \& Moncet(1992)}]{Clough1992}
Clough SA, Iacono MJ, Moncet JL. 1992.
Line-by-line calculations of atmospheric fluxes and cooling rates: Application
  to water vapor (paper 92jd01419).
\textit{Journal of Geophysical Research} 97:15--761

\bibitem[{Connerney et~al.(2001)Connerney, Acuna, Wasilewski, Kletetschka, Ness
  et~al.}]{Connerney2001}
Connerney J, Acuna M, Wasilewski P, Kletetschka G, Ness N, et~al. 2001.
The global magnetic field of mars and implications for crustal evolution.
\textit{Geophys. Res. Lett} 28:4015--4018

\bibitem[{Craddock \& Howard(2002)}]{Craddock2002}
Craddock RA, Howard AD. 2002.
The case for rainfall on a warm, wet early mars.
\textit{Journal of Geophysical Research} 107:5111

\bibitem[{Dauphas \& Pourmand(2011)}]{Dauphas2011}
Dauphas N, Pourmand A. 2011.
Hf-w-th evidence for rapid growth of mars and its status as a planetary embryo.
\textit{Nature} 473:489--492

\bibitem[{D'Hondt et~al.(2004)D'Hondt, J{\o}rgensen, Miller, Batzke, Blake
  et~al.}]{dHondt2004}
D'Hondt S, J{\o}rgensen BB, Miller DJ, Batzke A, Blake R, et~al. 2004.
Distributions of microbial activities in deep subseafloor sediments.
\textit{Science} 306:2216--2221

\bibitem[{{di Achille} \& {Hynek}(2010)}]{diAchille2010}
{di Achille} G, {Hynek} BM. 2010.
{Ancient ocean on Mars supported by global distribution of deltas and valleys}.
\textit{Nature Geoscience} 3:459--463

\bibitem[{Edwards \& Ehlmann(2015)}]{Edwards2015}
Edwards CS, Ehlmann BL. 2015.
Carbon sequestration on mars.
\textit{Geology} 43:863--866

\bibitem[{Ehlmann \& Dundar(2015)}]{Ehlmann2015_LPSC}
Ehlmann B, Dundar M. 2015.
In \textit{Lunar and Planetary Science Conference}, vol.~46

\bibitem[{Ehlmann \& Edwards(2014)}]{Ehlmann2014}
Ehlmann BL, Edwards CS. 2014.
Mineralogy of the martian surface.
\textit{Annual Review of Earth and Planetary Sciences} 42:291--315

\bibitem[{{Ehlmann} et~al.(2008){Ehlmann}, {Mustard}, {Fassett}, {Schon},
  {Head} et~al.}]{Ehlmann2008}
{Ehlmann} BL, {Mustard} JF, {Fassett} CI, {Schon} SC, {Head} III JW, et~al.
  2008.
{Clay minerals in delta deposits and organic preservation potential on Mars}.
\textit{Nature Geoscience} 1:355--358

\bibitem[{{Ehlmann} et~al.(2011){Ehlmann}, {Mustard}, {Murchie}, {Bibring},
  {Meunier} et~al.}]{Ehlmann2011}
{Ehlmann} BL, {Mustard} JF, {Murchie} SL, {Bibring} JP, {Meunier} A, et~al.
  2011.
{Subsurface water and clay mineral formation during the early history of Mars}.
\textit{Nature} 479:53--60

\bibitem[{Ehlmann et~al.(2009)Ehlmann, Mustard, Swayze, Clark, Bishop
  et~al.}]{Ehlmann2009}
Ehlmann BL, Mustard JF, Swayze GA, Clark RN, Bishop JL, et~al. 2009.
Identification of hydrated silicate minerals on mars using mro-crism: Geologic
  context near nili fossae and implications for aqueous alteration.
\textit{Journal of Geophysical Research: Planets (1991--2012)} 114

\bibitem[{{Fair{\'e}n}(2010)}]{Fairen2010}
{Fair{\'e}n} AG. 2010.
{A cold and wet Mars}.
\textit{Icarus} 208:165--175

\bibitem[{Fair{\'e}n et~al.(2009)Fair{\'e}n, Davila, Gago-Duport, Amils \&
  McKay}]{Fairen2009}
Fair{\'e}n AG, Davila AF, Gago-Duport L, Amils R, McKay CP. 2009.
Stability against freezing of aqueous solutions on early mars.
\textit{Nature} 459:401--404

\bibitem[{{Fassett} \& {Head}(2008)}]{Fassett2008}
{Fassett} CI, {Head} JW. 2008.
{The timing of martian valley network activity: Constraints from buffered
  crater counting}.
\textit{Icarus} 195:61--89

\bibitem[{{Fassett} \& {Head}(2011)}]{Fassett2011}
{Fassett} CI, {Head} JW. 2011.
{Sequence and timing of conditions on early Mars}.
\textit{Icarus} 211:1204--1214

\bibitem[{Fastook \& Head(2014)}]{Fastook2015}
Fastook JL, Head JW. 2014.
Glaciation in the late noachian icy highlands: Ice accumulation, distribution,
  flow rates, basal melting, and top-down melting rates and patterns.
\textit{Planetary and Space Science}

\bibitem[{Fastook et~al.(2012)Fastook, Head, Marchant, Forget \&
  Madeleine}]{Fastook2012}
Fastook JL, Head JW, Marchant DR, Forget F, Madeleine JB. 2012.
Early mars climate near the noachian{\~n}hesperian boundary: Independent
  evidence for cold conditions from basal melting of the south polar ice sheet
  (dorsa argentea formation) and implications for valley network formation.
\textit{Icarus} 219:25 -- 40

\bibitem[{{Forget} et~al.(2006){Forget}, {Haberle}, {Montmessin}, {Levrard} \&
  {Head}}]{Forget2006}
{Forget} F, {Haberle} RM, {Montmessin} F, {Levrard} B, {Head} JW. 2006.
{Formation of Glaciers on Mars by Atmospheric Precipitation at High Obliquity}.
\textit{Science} 311:368--371

\bibitem[{{Forget} \& {Pierrehumbert}(1997)}]{Forget1997}
{Forget} F, {Pierrehumbert} RT. 1997.
{Warming Early Mars with Carbon Dioxide Clouds That Scatter Infrared
  Radiation}.
\textit{Science} 278:1273--+

\bibitem[{Forget et~al.(2013)Forget, Wordsworth, Millour, Madeleine, Kerber
  et~al.}]{Forget2013}
Forget F, Wordsworth RD, Millour E, Madeleine JB, Kerber L, et~al. 2013.
3d modelling of the early martian climate under a denser {CO2} atmosphere:
  Temperatures and {CO2} ice clouds.
\textit{Icarus}

\bibitem[{Forster et~al.(2007)Forster, Ramaswamy, Artaxo, Berntsen, Betts
  et~al.}]{Forster2007}
Forster P, Ramaswamy V, Artaxo P, Berntsen T, Betts R, et~al. 2007.
In \textit{Climate Change 2007. The Physical Science Basis}. IPCC

\bibitem[{Gendrin et~al.(2005)Gendrin, Mangold, Bibring, Langevin, Gondet
  et~al.}]{Gendrin2005}
Gendrin A, Mangold N, Bibring JP, Langevin Y, Gondet B, et~al. 2005.
Sulfates in martian layered terrains: the omega/mars express view.
\textit{Science} 307:1587--1591

\bibitem[{Goody et~al.(1989)Goody, West, Chen \& Crisp}]{Goody1989}
Goody R, West R, Chen L, Crisp D. 1989.
The correlated-k method for radiation calculations in nonhomogeneous
  atmospheres.
\textit{Journal of Quantitative Spectroscopy and Radiative Transfer}
  42:539--550

\bibitem[{{Goudge} et~al.(2012){Goudge}, {Mustard}, {Head} \&
  {Fassett}}]{Goudge2012}
{Goudge} TA, {Mustard} JF, {Head} JW, {Fassett} CI. 2012.
In \textit{Lunar and Planetary Institute Science Conference Abstracts}, vol.~43
  of \textit{Lunar and Planetary Institute Science Conference Abstracts}

\bibitem[{{Gough}(1981)}]{Gough1981}
{Gough} DO. 1981.
{Solar interior structure and luminosity variations}.
\textit{Solar Physics} 74:21--34

\bibitem[{Greenwood et~al.(2008)Greenwood, Itoh, Sakamoto, Vicenzi \&
  Yurimoto}]{Greenwood2008}
Greenwood JP, Itoh S, Sakamoto N, Vicenzi EP, Yurimoto H. 2008.
Hydrogen isotope evidence for loss of water from mars through time.
\textit{Geophysical Research Letters} 35

\bibitem[{{Grott} et~al.(2011){Grott}, {Morschhauser}, {Breuer} \&
  {Hauber}}]{Grott2011}
{Grott} M, {Morschhauser} A, {Breuer} D, {Hauber} E. 2011.
{Volcanic outgassing of CO$_{2}$ and H$_{2}$O on Mars}.
\textit{Earth and Planetary Science Letters} 308:391--400

\bibitem[{Grotzinger et~al.(2014)Grotzinger, Sumner, Kah, Stack, Gupta
  et~al.}]{Grotzinger2014}
Grotzinger JP, Sumner DY, Kah LC, Stack K, Gupta S, et~al. 2014.
A habitable fluvio-lacustrine environment at yellowknife bay, gale crater,
  mars.
\textit{Science} 343

\bibitem[{{Gruszka} \& {Borysow}(1997)}]{Gruszka1997}
{Gruszka} M, {Borysow} A. 1997.
{Roto-Translational Collision-Induced Absorption of {CO2} for the Atmosphere of
  Venus at Frequencies from 0 to 250 cm\^{}-1, at Temperatures from 200 to 800
  K}.
\textit{Icarus} 129:172--177

\bibitem[{{Haberle}(1998)}]{Haberle1998}
{Haberle} RM. 1998.
{Early Mars climate models}.
\textit{J.~Geophys.~Res.} 103:28467

\bibitem[{Halevy \& {Head}(2014)}]{Halevy2014}
Halevy I, {Head} JW. 2014.
Episodic warming of early mars by punctuated volcanism.
\textit{Nature Geoscience}

\bibitem[{Halevy, Pierrehumbert \& Schrag(2009)}]{Halevy2009}
Halevy I, Pierrehumbert R, Schrag D. 2009.
Radiative transfer in co2-rich paleoatmospheres.
\textit{Journal of Geophysical Research: Atmospheres (1984--2012)} 114

\bibitem[{{Halevy}, {Zuber} \& {Schrag}(2007)}]{Halevy2007}
{Halevy} I, {Zuber} MT, {Schrag} DP. 2007.
{A sulfur dioxide climate feedback on early Mars}.
\textit{Science} 318:1903--

\bibitem[{Hartmann \& Neukum(2001)}]{Hartmann2001}
Hartmann WK, Neukum G. 2001.
In \textit{Chronology and evolution of Mars}. Springer,  165--194

\bibitem[{Hayes \& Waldbauer(2006)}]{Hayes2006}
Hayes JM, Waldbauer JR. 2006.
The carbon cycle and associated redox processes through time.
\textit{Philosophical Transactions of the Royal Society B: Biological Sciences}
  361:931--950

\bibitem[{{Head}, {Kreslavsky} \& {Pratt}(2002)}]{Head2002}
{Head} JW, {Kreslavsky} MA, {Pratt} S. 2002.
{Northern lowlands of Mars: Evidence for widespread volcanic flooding and
  tectonic deformation in the Hesperian Period}.
\textit{Journal of Geophysical Research (Planets)} 107:5003

\bibitem[{Head \& Marchant(2014)}]{Head2014}
Head JW, Marchant DR. 2014.
The climate history of early mars: insights from the antarctic mcmurdo dry
  valleys hydrologic system.
\textit{Antarctic Science} 26:774--800

\bibitem[{{Head} \& {Pratt}(2001)}]{HeadPratt2001}
{Head} JW, {Pratt} S. 2001.
{Extensive Hesperian-aged south polar ice sheet on Mars: Evidence for massive
  melting and retreat, and lateral flow and ponding of meltwater}.
\textit{Journal of Geophysical Research} 106:12275--12300

\bibitem[{{Head} et~al.(1999){Head}, {Hiesinger}, {Ivanov}, {Kreslavsky},
  {Pratt} \& {Thomson}}]{Head1999}
{Head} III JW, {Hiesinger} H, {Ivanov} MA, {Kreslavsky} MA, {Pratt} S,
  {Thomson} BJ. 1999.
{Possible ancient oceans on Mars: evidence from Mars Orbiter Laser Altimeter
  data.}
\textit{Science} 286:2134--2137

\bibitem[{Hirschmann \& Withers(2008)}]{HirschmannWithers2008}
Hirschmann MM, Withers AC. 2008.
Ventilation of {CO2} from a reduced mantle and consequences for the early
  martian greenhouse.
\textit{Earth and Planetary Science Letters} 270:147--155

\bibitem[{Hoffman et~al.(1998)Hoffman, Kaufman, Halverson \&
  Schrag}]{Hoffman1998}
Hoffman PF, Kaufman AJ, Halverson GP, Schrag DP. 1998.
A neoproterozoic snowball earth.
\textit{science} 281:1342--1346

\bibitem[{{Hoke}, {Hynek} \& Tucker(2011)}]{Hoke2011}
{Hoke} MRT, {Hynek} BM, Tucker GE. 2011.
Formation timescales of large martian valley networks.
\textit{Earth and Planetary Science Letters} 312:1--12

\bibitem[{{Howard}(1981)}]{Howard1981}
{Howard} AD. 1981.
Etched plains and braided ridges of the south polar region of mars: Features
  produced by basal melting of ground ice?
Reports of Planetary Geology Program 84211, NASA

\bibitem[{Howard, Moore \& Irwin(2005)}]{Howard2005}
Howard AD, Moore JM, Irwin RP. 2005.
An intense terminal epoch of widespread fluvial activity on early mars: 1.
  valley network incision and associated deposits.
\textit{Journal of Geophysical Research: Planets (1991--2012)} 110

\bibitem[{{Hynek}, {Beach} \& {Hoke}(2010)}]{Hynek2010}
{Hynek} BM, {Beach} M, {Hoke} MRT. 2010.
{Updated global map of Martian valley networks and implications for climate and
  hydrologic processes}.
\textit{Journal of Geophysical Research (Planets)} 115:E09008

\bibitem[{{Irwin} et~al.(2005){Irwin}, {Howard}, {Craddock} \&
  {Moore}}]{Irwin2005}
{Irwin} RP, {Howard} AD, {Craddock} RA, {Moore} JM. 2005.
{An intense terminal epoch of widespread fluvial activity on early Mars: 2.
  Increased runoff and paleolake development}.
\textit{Journal of Geophysical Research (Planets)} 110:12--+

\bibitem[{{Jakosky} \& Jones(1997)}]{Jakosky1997}
{Jakosky} BM, Jones JH. 1997.
The history of martian volatiles.
\textit{Reviews of Geophysics} 35:1--16

\bibitem[{{Johnson} et~al.(2008){Johnson}, {Mischna}, {Grove} \&
  {Zuber}}]{StewartJ2008}
{Johnson} SS, {Mischna} MA, {Grove} TL, {Zuber} MT. 2008.
{Sulfur-induced greenhouse warming on early Mars}.
\textit{Journal of Geophysical Research (Planets)} 113:8005--+

\bibitem[{{Johnson}, {Pavlov} \& {Mischna}(2009)}]{Johnson2009}
{Johnson} SSS, {Pavlov} AA, {Mischna} MA. 2009.
Fate of so2 in the ancient martian atmosphere: Implications for transient
  greenhouse warming.
\textit{Journal of Geophysical Research: Planets (1991--2012)} 114

\bibitem[{Kahre et~al.(2013)Kahre, Vines, Haberle \& Hollingsworth}]{Kahre2013}
Kahre MA, Vines SK, Haberle RM, Hollingsworth JL. 2013.
The early martian atmosphere: Investigating the role of the dust cycle in the
  possible maintenance of two stable climate states.
\textit{Journal of Geophysical Research: Planets} 118:1388--1396

\bibitem[{Kargel \& Strom(1992)}]{Kargel1992}
Kargel JS, Strom RG. 1992.
Ancient glaciation on mars.
\textit{Geology} 20:3--7

\bibitem[{{Kasting}(1991)}]{Kasting1991}
{Kasting} JF. 1991.
{CO2 condensation and the climate of early Mars}.
\textit{Icarus} 94:1--13

\bibitem[{Kasting(1997)}]{Kasting1997}
Kasting JF. 1997.
Warming early earth and mars.
\textit{Science} 276:1213

\bibitem[{{Kasting}, {Whitmire} \& {Reynolds}(1993)}]{Kasting1993}
{Kasting} JF, {Whitmire} DP, {Reynolds} RT. 1993.
{Habitable Zones around Main Sequence Stars}.
\textit{Icarus} 101:108--128

\bibitem[{{Kerber}, {Forget} \& {Wordsworth}(2015)}]{Kerber2015}
{Kerber} L, {Forget} F, {Wordsworth} RD. 2015.
Sulfur in the early martian atmosphere revisited: Experiments with a 3-d global
  climate model.
\textit{Icarus}

\bibitem[{Khairoutdinov \& Randall(2001)}]{Khairoutdinov2001}
Khairoutdinov MF, Randall DA. 2001.
A cloud resolving model as a cloud parameterization in the ncar community
  climate system model: Preliminary results.
\textit{Geophysical Research Letters} 28:3617--3620

\bibitem[{Kirschvink(1992)}]{Kirschvink1992}
Kirschvink JL. 1992.
Late proterozoic low-latitude global glaciation: the snowball earth.
Cambridge University Press

\bibitem[{Kirschvink \& Weiss(2002)}]{Kirschvink2002}
Kirschvink JL, Weiss BP. 2002.
Mars, panspermia, and the origin of life: where did it all begin.
\textit{Palaeontologia Electronica} 4:8--15

\bibitem[{Kite et~al.(2013)Kite, Halevy, Kahre, Wolff \& Manga}]{Kite2013}
Kite ES, Halevy I, Kahre MA, Wolff MJ, Manga M. 2013.
Seasonal melting and the formation of sedimentary rocks on mars, with
  predictions for the gale crater mound.
\textit{Icarus} 223:181--210

\bibitem[{Kite et~al.(2014)Kite, Williams, Lucas \& Aharonson}]{Kite2014}
Kite ES, Williams JP, Lucas A, Aharonson O. 2014.
Low palaeopressure of the martian atmosphere estimated from the size
  distribution of ancient craters.
\textit{Nature Geoscience} 7:335--339

\bibitem[{{Kitzmann}, {Patzer} \& {Rauer}(2013)}]{Kitzmann2013}
{Kitzmann} D, {Patzer} ABC, {Rauer} H. 2013.
{Clouds in the atmospheres of extrasolar planets. IV. On the scattering
  greenhouse effect of CO$_{2}$ ice particles: Numerical radiative transfer
  studies}.
\textit{Astronomy \& Astrophysics} 557:A6

\bibitem[{Lacis \& Oinas(1991)}]{Lacis1991}
Lacis AA, Oinas V. 1991.
A description of the correlated k distribution method for modeling nongray
  gaseous absorption, thermal emission, and multiple scattering in vertically
  inhomogeneous atmospheres.
\textit{J. geophys. Res} 96:9027--9064

\bibitem[{Lammer et~al.(2013)Lammer, Chassefi{\'R}re, Karatekin, Morschhauser,
  Niles et~al.}]{Lammer2013}
Lammer H, Chassefi{\'R}re E, Karatekin {\'I}, Morschhauser A, Niles PB, et~al.
  2013.
Outgassing history and escape of the martian atmosphere and water inventory.
\textit{Space Science Reviews} 174:113--154

\bibitem[{{Lammer} et~al.(2003){Lammer}, {Selsis}, {Ribas}, {Guinan}, {Bauer}
  \& {Weiss}}]{Lammer2003}
{Lammer} H, {Selsis} F, {Ribas} I, {Guinan} EF, {Bauer} SJ, {Weiss} WW. 2003.
{Atmospheric Loss of Exoplanets Resulting from Stellar X-Ray and
  Extreme-Ultraviolet Heating}.
\textit{The Astrophysical Journal Letters} 598:L121--L124

\bibitem[{{Laskar} et~al.(2004){Laskar}, {Correia}, {Gastineau}, {Joutel},
  {Levrard} \& {Robutel}}]{Laskar2004}
{Laskar} J, {Correia} ACM, {Gastineau} M, {Joutel} F, {Levrard} B, {Robutel} P.
  2004.
{Long term evolution and chaotic diffusion of the insolation quantities of
  Mars}.
\textit{Icarus} 170:343--364

\bibitem[{Laskar \& Robutel(1993)}]{Laskar1993}
Laskar J, Robutel P. 1993.
The chaotic obliquity of the planets.
\textit{Nature} 361:608--612

\bibitem[{Leovy \& Mintz(1969)}]{Leovy1969}
Leovy C, Mintz Y. 1969.
Numerical simulation of the atmospheric circulation and climate of mars.
\textit{Journal of the Atmospheric Sciences} 26:1167--1190

\bibitem[{{Madeleine} et~al.(2009){Madeleine}, {Forget}, {Head}, {Levrard},
  {Montmessin} \& {Millour}}]{Madeleine2009}
{Madeleine} JB, {Forget} F, {Head} JW, {Levrard} B, {Montmessin} F, {Millour}
  E. 2009.
{Amazonian northern mid-latitude glaciation on Mars: A proposed climate
  scenario}.
\textit{Icarus} 203:390--405

\bibitem[{{Malin} \& {Edgett}(2003)}]{Malin2003}
{Malin} MC, {Edgett} KS. 2003.
{Evidence for Persistent Flow and Aqueous Sedimentation on Early Mars}.
\textit{Science} 302:1931--1934

\bibitem[{{Malin} \& {Edgett}(1999)}]{Malin1999}
{Malin} MCC, {Edgett} KS. 1999.
Oceans or seas in the martian northern lowlands: High resolution imaging tests
  of proposed coastlines.
\textit{Geophysical Research Letters} 26:3049--3052

\bibitem[{Mangold et~al.(2004)Mangold, Quantin, Ansan, Delacourt \&
  Allemand}]{Mangold2004}
Mangold N, Quantin C, Ansan V, Delacourt C, Allemand P. 2004.
Evidence for precipitation on mars from dendritic valleys in the valles
  marineris area.
\textit{Science} 305:78--81

\bibitem[{Marty(2012)}]{Marty2012}
Marty B. 2012.
The origins and concentrations of water, carbon, nitrogen and noble gases on
  earth.
\textit{Earth and Planetary Science Letters} 313:56--66

\bibitem[{Matsubara, Howard \& Gochenour(2013)}]{Matsubara2013}
Matsubara Y, Howard AD, Gochenour JP. 2013.
Hydrology of early mars: Valley network incision.
\textit{Journal of Geophysical Research: Planets} 118:1365--1387

\bibitem[{{Michalski} \& {Niles}(2010)}]{Michalski2010}
{Michalski} JR, {Niles} PB. 2010.
{Deep crustal carbonate rocks exposed by meteor impact on Mars}.
\textit{Nature Geoscience} 3:751--755

\bibitem[{Mileikowsky et~al.(2000)Mileikowsky, Cucinotta, Wilson, Gladman,
  Horneck et~al.}]{Mileikowsky2000}
Mileikowsky C, Cucinotta FA, Wilson JW, Gladman B, Horneck G, et~al. 2000.
Natural transfer of viable microbes in space: 1. from mars to earth and earth
  to mars.
\textit{Icarus} 145:391--427

\bibitem[{Milton(1973)}]{Milton1973}
Milton DJ. 1973.
Water and processes of degradation in the martian landscape.
\textit{Journal of Geophysical Research} 78:4037--4047

\bibitem[{Minton \& Malhotra(2007)}]{Minton2007}
Minton DA, Malhotra R. 2007.
Assessing the massive young sun hypothesis to solve the warm young earth
  puzzle.
\textit{The Astrophysical Journal} 660:1700

\bibitem[{Mischna et~al.(2013)Mischna, Baker, Milliken, Richardson \&
  Lee}]{Mischna2013}
Mischna MA, Baker V, Milliken R, Richardson M, Lee C. 2013.
Effects of obliquity and water vapor/trace gas greenhouses in the early martian
  climate.
\textit{Journal of Geophysical Research: Planets}

\bibitem[{Mischna et~al.(2003)Mischna, Richardson, Wilson \&
  McCleese}]{Mischna2003}
Mischna MA, Richardson MI, Wilson RJ, McCleese DJ. 2003.
On the orbital forcing of martian water and {CO2} cycles: A general circulation
  model study with simplified volatile schemes.
\textit{Journal of Geophysical Research: Planets (1991--2012)} 108

\bibitem[{{Montmessin} et~al.(2007){Montmessin}, {Gondet}, {Bibring},
  {Langevin}, {Drossart} et~al.}]{Montmessin2007}
{Montmessin} F, {Gondet} B, {Bibring} J, {Langevin} Y, {Drossart} P, et~al.
  2007.
{Hyperspectral imaging of convective CO2 ice clouds in the equatorial
  mesosphere of Mars}.
\textit{JGR (Planets)} 112:11--+

\bibitem[{{Morris} et~al.(2010){Morris}, {Ruff}, {Gellert}, {Ming}, {Arvidson}
  et~al.}]{Morris2010}
{Morris} RV, {Ruff} SW, {Gellert} R, {Ming} DW, {Arvidson} RE, et~al. 2010.
{Identification of Carbonate-Rich Outcrops on Mars by the Spirit Rover}.
\textit{Science} 329:421--

\bibitem[{{Murchie} et~al.(2009){Murchie}, {Mustard}, {Ehlmann}, {Milliken},
  {Bishop} et~al.}]{Murchie2009}
{Murchie} SL, {Mustard} JF, {Ehlmann} BL, {Milliken} RE, {Bishop} JL, et~al.
  2009.
{A synthesis of Martian aqueous mineralogy after 1 Mars year of observations
  from the Mars Reconnaissance Orbiter}.
\textit{Journal of Geophysical Research (Planets)} 114:0

\bibitem[{{Mustard} et~al.(2008){Mustard}, {Murchie}, {Pelkey}, {Ehlmann},
  {Milliken} et~al.}]{Mustard2008}
{Mustard} JF, {Murchie} SL, {Pelkey} SM, {Ehlmann} BL, {Milliken} RE, et~al.
  2008.
{Hydrated silicate minerals on Mars observed by the Mars Reconnaissance Orbiter
  CRISM instrument}.
\textit{Nature} 454:305--309

\bibitem[{Nakajima, Hayashi \& Abe(1992)}]{Nakajima1992}
Nakajima S, Hayashi YY, Abe Y. 1992.
A study on the {\"e}runaway greenhouse effect{\'\i}with a one-dimensional
  radiative--convective equilibrium model.
\textit{J. Atmos. Sci} 49:2256--2266

\bibitem[{Niles et~al.(2013)Niles, Catling, Berger, Chassefi{\`e}re, Ehlmann
  et~al.}]{Niles2013}
Niles PB, Catling DC, Berger G, Chassefi{\`e}re E, Ehlmann BL, et~al. 2013.
Geochemistry of carbonates on mars: implications for climate history and nature
  of aqueous environments.
\textit{Space Science Reviews} 174:301--328

\bibitem[{Nimmo \& Tanaka(2005)}]{Nimmo2005}
Nimmo F, Tanaka K. 2005.
Early crustal evolution of mars.
\textit{Annu. Rev. Earth Planet. Sci.} 33:133--161

\bibitem[{Osterloo et~al.(2010)Osterloo, Anderson, Hamilton \&
  Hynek}]{Osterloo2010}
Osterloo MM, Anderson FS, Hamilton VE, Hynek BM. 2010.
Geologic context of proposed chloride-bearing materials on mars.
\textit{Journal of Geophysical Research: Planets (1991--2012)} 115

\bibitem[{{Parker} et~al.(1993){Parker}, Gorsline, Saunders, Pieri \&
  Schneeberger}]{Parker1993}
{Parker} TJJ, Gorsline DS, Saunders RS, Pieri DC, Schneeberger DM. 1993.
Coastal geomorphology of the martian northern plains.
\textit{Journal of Geophysical Research: Planets (1991--2012)} 98:11061--11078

\bibitem[{{Perrin} \& {Hartmann}(1989)}]{Perrin1989}
{Perrin} MY, {Hartmann} JM. 1989.
{Temperature-dependent measurements and modeling of absorption by CO2-N2
  mixtures in the far line-wings of the 4.3-micron CO2 band}.
\textit{J. Quant. Spectrosc. Radiat. Transfer} 42:311--317

\bibitem[{Perron et~al.(2007)Perron, Mitrovica, Manga, Matsuyama \&
  Richards}]{Perron2007}
Perron JT, Mitrovica JX, Manga M, Matsuyama I, Richards MA. 2007.
Evidence for an ancient martian ocean in the topography of deformed shorelines.
\textit{Nature} 447:840--843

\bibitem[{{Phillips} et~al.(2001){Phillips}, {Zuber}, {Solomon}, {Golombek},
  {Jakosky} et~al.}]{Phillips2001}
{Phillips} RJ, {Zuber} MT, {Solomon} SC, {Golombek} MP, {Jakosky} BM, et~al.
  2001.
{Ancient geodynamics and global-scale hydrology on Mars}.
\textit{Science} 291:2587--2591

\bibitem[{{Pierrehumbert} et~al.(2011){Pierrehumbert}, Abbot, Voigt \&
  Koll}]{Pierrehumbert2011_NP}
{Pierrehumbert} RT, Abbot DS, Voigt A, Koll D. 2011.
Climate of the neoproterozoic.
\textit{Annual Review of Earth and Planetary Sciences} 39:417

\bibitem[{{Pollack} et~al.(1987){Pollack}, {Kasting}, {Richardson} \&
  {Poliakoff}}]{Pollack1987}
{Pollack} JB, {Kasting} JF, {Richardson} SM, {Poliakoff} K. 1987.
{The case for a wet, warm climate on early Mars}.
\textit{Icarus} 71:203--224

\bibitem[{{Postawko} \& {Kuhn}(1986)}]{Postawko1986}
{Postawko} SE, {Kuhn} WR. 1986.
{Effect of the greenhouse gases (CO$_{2}$, H$_{2}$O, SO$_{2}$) on martian
  paleoclimate}.
\textit{Journal of Geophysical Research} 91:431--D438

\bibitem[{{Poulet} et~al.(2005){Poulet}, {Bibring}, {Mustard}, {Gendrin},
  {Mangold} et~al.}]{Poulet2005}
{Poulet} F, {Bibring} JP, {Mustard} JF, {Gendrin} A, {Mangold} N, et~al. 2005.
{Phyllosilicates on Mars and implications for early martian climate}.
\textit{Nature} 438:623--627

\bibitem[{{Quintana} et~al.(2014){Quintana}, Barclay, Raymond, Rowe, Bolmont
  et~al.}]{Quintana2014}
{Quintana} EVV, Barclay T, Raymond SN, Rowe JF, Bolmont E, et~al. 2014.
An earth-sized planet in the habitable zone of a cool star.
\textit{Science} 344:277--280

\bibitem[{Ramirez et~al.(2014)Ramirez, Kopparapu, Zugger, Robinson, Freedman \&
  Kasting}]{Ramirez2014}
Ramirez RM, Kopparapu R, Zugger ME, Robinson TD, Freedman R, Kasting JF. 2014.
Warming early mars with {CO2} and {H2}.
\textit{Nature Geoscience} 7:59--63

\bibitem[{Richardson \& Mischna(2005)}]{Richardson2005}
Richardson MI, Mischna MA. 2005.
Long-term evolution of transient liquid water on mars.
\textit{Journal of Geophysical Research: Planets (1991--2012)} 110

\bibitem[{{Sagan}(1977)}]{Sagan1977}
{Sagan} C. 1977.
Reducing greenhouses and the temperature history of earth and mars.
\textit{Nature}

\bibitem[{Scanlon et~al.(2013)Scanlon, Head, Madeleine, Wordsworth \&
  Forget}]{Scanlon2013}
Scanlon KE, Head JW, Madeleine JB, Wordsworth RD, Forget F. 2013.
Orographic precipitation in valley network headwaters: Constraints on the
  ancient martian atmosphere.
\textit{Geophysical Research Letters} 40:4182--4187

\bibitem[{Scott \& Tanaka(1986)}]{Scott1986}
Scott DH, Tanaka KL. 1986.
Geologic map of the western equatorial region of mars.
Geological Survey (US)

\bibitem[{Segura, McKay \& Toon(2012)}]{Segura2012}
Segura TL, McKay CP, Toon OB. 2012.
An impact-induced, stable, runaway climate on mars.
\textit{Icarus} 220:144--148

\bibitem[{{Segura}, {Toon} \& {Colaprete}(2008)}]{Segura2008}
{Segura} TL, {Toon} OB, {Colaprete} A. 2008.
{Modeling the environmental effects of moderate-sized impacts on Mars}.
\textit{Journal of Geophysical Research (Planets)} 113:E11007

\bibitem[{{Segura} et~al.(2002){Segura}, {Toon}, {Colaprete} \&
  {Zahnle}}]{Segura2002}
{Segura} TL, {Toon} OB, {Colaprete} A, {Zahnle} K. 2002.
{Environmental Effects of Large Impacts on Mars}.
\textit{Science} 298:1977--1980

\bibitem[{{Sleep} \& {Zahnle}(2001)}]{Sleep2001}
{Sleep} NH, {Zahnle} K. 2001.
{Carbon dioxide cycling and implications for climate on ancient Earth}.
\textit{Journal of Geophysical Research} 106:1373--1400

\bibitem[{Solomon et~al.(2005)Solomon, Aharonson, Aurnou, Banerdt, Carr
  et~al.}]{Solomon2005}
Solomon SC, Aharonson O, Aurnou JM, Banerdt WB, Carr MH, et~al. 2005.
New perspectives on ancient mars.
\textit{Science} 307:1214--1220

\bibitem[{Soto et~al.(2015)Soto, Mischna, Schneider, Lee \&
  Richardson}]{Soto2015}
Soto A, Mischna M, Schneider T, Lee C, Richardson M. 2015.
Martian atmospheric collapse: Idealized gcm studies.
\textit{Icarus} 250:553--569

\bibitem[{{Squyres} \& {Kasting}(1994)}]{Squyres1994}
{Squyres} SW, {Kasting} JF. 1994.
{Early Mars: How Warm and How Wet?}
\textit{Science} 265:744--749

\bibitem[{Stenchikov et~al.(1998)Stenchikov, Kirchner, Robock, Graf, Antuna
  et~al.}]{Stenchikov1998}
Stenchikov GL, Kirchner I, Robock A, Graf HF, Antuna JC, et~al. 1998.
Radiative forcing from the 1991 mount pinatubo volcanic eruption.
\textit{Journal of Geophysical Research: Atmospheres (1984--2012)}
  103:13837--13857

\bibitem[{Stepinski \& Stepinski(2005)}]{Stepinski2005}
Stepinski TF, Stepinski AP. 2005.
Morphology of drainage basins as an indicator of climate on early mars.
\textit{Journal of Geophysical Research: Planets (1991--2012)} 110

\bibitem[{Stroeve et~al.(2007)Stroeve, Holland, Meier, Scambos \&
  Serreze}]{Stroeve2007}
Stroeve J, Holland MM, Meier W, Scambos T, Serreze M. 2007.
Arctic sea ice decline: Faster than forecast.
\textit{Geophysical research letters} 34

\bibitem[{Tanaka(1986)}]{Tanaka1986}
Tanaka KL. 1986.
The stratigraphy of mars.
\textit{Journal of Geophysical Research: Solid Earth (1978--2012)}
  91:E139--E158

\bibitem[{Tanaka \& Scott(1987)}]{Tanaka1987}
Tanaka KL, Scott DH. 1987.
Geologic map of the polar regions of mars.
Geological Survey (US)

\bibitem[{Tanaka et~al.(2014)Tanaka, Skinner, Dohm, Irwin~III, Kolb
  et~al.}]{Tanaka2014}
Tanaka KL, Skinner JA, Dohm JM, Irwin~III RP, Kolb EJ, et~al. 2014.
Geologic map of mars.
US Department of the Interior, US Geological Survey

\bibitem[{{Tian} et~al.(2010){Tian}, {Claire}, {Haqq-Misra}, {Smith}, {Crisp}
  et~al.}]{Tian2010}
{Tian} F, {Claire} MW, {Haqq-Misra} JD, {Smith} M, {Crisp} DC, et~al. 2010.
{Photochemical and climate consequences of sulfur outgassing on early Mars}.
\textit{Earth and Planetary Science Letters} 295:412--418

\bibitem[{{Toon}, {Segura} \& {Zahnle}(2010)}]{Toon2010}
{Toon} OB, {Segura} T, {Zahnle} K. 2010.
{The formation of martian river valleys by impacts}.
\textit{Annual Review of Earth and Planetary Sciences} 38:303--322

\bibitem[{Tosca \& Knoll(2009)}]{Tosca2009}
Tosca NJ, Knoll AH. 2009.
Juvenile chemical sediments and the long term persistence of water at the
  surface of mars.
\textit{Earth and Planetary Science Letters} 286:379--386

\bibitem[{{Udry} et~al.(2007){Udry}, {Bonfils}, {Delfosse}, {Forveille},
  {Mayor} et~al.}]{Udry2007}
{Udry} S, {Bonfils} X, {Delfosse} X, {Forveille} T, {Mayor} M, et~al. 2007.
{The HARPS search for southern extra-solar planets. XI. Super-Earths (5 and 8
  M+) in a 3-planet system}.
\textit{Astron. Astrophys.} 469:L43--L47

\bibitem[{Urata \& Toon(2013)}]{Urata2013}
Urata RA, Toon OB. 2013.
Simulations of the martian hydrologic cycle with a general circulation model:
  Implications for the ancient martian climate.
\textit{Icarus}

\bibitem[{Villanueva et~al.(2015)Villanueva, Mumma, Novak, K{\"a}ufl, Hartogh
  et~al.}]{Villanueva2015}
Villanueva G, Mumma M, Novak R, K{\"a}ufl H, Hartogh P, et~al. 2015.
Strong water isotopic anomalies in the martian atmosphere: Probing current and
  ancient reservoirs.
\textit{Science} 348:218--221

\bibitem[{von Paris et~al.(2013)von Paris, {Grenfell}, {Rauer} \&
  Stock}]{von2013}
von Paris P, {Grenfell} JLL, {Rauer} H, Stock JW. 2013.
N2-associated surface warming on early mars.
\textit{Planetary and Space Science} 82:149--154

\bibitem[{Wadhwa(2001)}]{Wadhwa2001}
Wadhwa M. 2001.
Redox state of mars' upper mantle and crust from eu anomalies in shergottite
  pyroxenes.
\textit{Science} 291:1527--1530

\bibitem[{{Walker}, {Hayes} \& {Kasting}(1981)}]{Walker1981}
{Walker} JCG, {Hayes} PB, {Kasting} JF. 1981.
{A negative feedback mechanism for the long-term stabilization of the earth's
  surface temperature}.
\textit{Journal of Geophysical Research} 86:9776--9782

\bibitem[{Walsh et~al.(2011)Walsh, Morbidelli, Raymond, O'Brien \&
  Mandell}]{Walsh2011}
Walsh KJ, Morbidelli A, Raymond SN, O'Brien DP, Mandell AM. 2011.
A low mass for mars from jupiter/'s early gas-driven migration.
\textit{Nature} 475:206--209

\bibitem[{Webster et~al.(2013)Webster, Mahaffy, Flesch, Niles, Jones
  et~al.}]{Webster2013}
Webster CR, Mahaffy PR, Flesch GJ, Niles PB, Jones JH, et~al. 2013.
Isotope ratios of h, c, and o in co2 and h2o of the martian atmosphere.
\textit{Science} 341:260--263

\bibitem[{Werner \& {Tanaka}(2011)}]{Werner2011}
Werner SC, {Tanaka} KL. 2011.
Redefinition of the crater-density and absolute-age boundaries for the
  chronostratigraphic system of mars.
\textit{Icarus} 215:603--607

\bibitem[{Williams et~al.(2013)Williams, Grotzinger, Dietrich, Gupta, Sumner
  et~al.}]{Williams2013}
Williams RME, Grotzinger JP, Dietrich WE, Gupta S, Sumner DY, et~al. 2013.
Martian fluvial conglomerates at gale crater.
\textit{Science} 340:1068--1072

\bibitem[{{Wordsworth}, {Forget} \& {Eymet}(2010)}]{Wordsworth2010}
{Wordsworth} R, {Forget} F, {Eymet} V. 2010.
{Infrared collision-induced and far-line absorption in dense CO2 atmospheres}.
\textit{Icarus} 210:992--997

\bibitem[{Wordsworth et~al.(2013)Wordsworth, Forget, Millour, Head, Madeleine
  \& Charnay}]{Wordsworth2013a}
Wordsworth R, Forget F, Millour E, Head JW, Madeleine JB, Charnay B. 2013.
Global modelling of the early martian climate under a denser {CO2} atmosphere:
  Water cycle and ice evolution.
\textit{Icarus} 222:1--19

\bibitem[{Wordsworth et~al.(2015)Wordsworth, {Kerber}, Pierrehumbert, {Forget}
  \& {Head}}]{Wordsworth2015}
Wordsworth R, {Kerber} L, Pierrehumbert R, {Forget} F, {Head} III JW. 2015.
Comparison of ``warm and wet'' and ``cold and icy'' scenarios for early mars in
  a 3d climate model.
\textit{Journal of Geophysical Research (Planets)} In Press

\bibitem[{Wordsworth \& Pierrehumbert(2013)}]{Wordsworth2013c}
Wordsworth R, Pierrehumbert R. 2013.
Hydrogen-nitrogen greenhouse warming in earth's early atmosphere.
\textit{Science} 339:64--67

\bibitem[{Wray et~al.(2008)Wray, Ehlmann, Squyres, Mustard \& Kirk}]{Wray2008}
Wray J, Ehlmann B, Squyres S, Mustard J, Kirk R. 2008.
Compositional stratigraphy of clay-bearing layered deposits at mawrth vallis,
  mars.
\textit{Geophysical Research Letters} 35

\bibitem[{Yung, Nair \& Gerstell(1997)}]{Yung1997}
Yung YL, Nair H, Gerstell MF. 1997.
Co2 greenhouse in the early martian atmosphere: So2 inhibits condensation.
\textit{Icarus} 130:222--224

\end{thebibliography}
\end{document}